\documentclass[fleqn, usenatbib]{mnras}
\usepackage{amsmath}	
\usepackage{mathptmx}
\usepackage{txfonts}

\usepackage[T1]{fontenc}

\DeclareRobustCommand{\VAN}[3]{#2}
\let\VANthebibliography\thebibliography
\def\thebibliography{\DeclareRobustCommand{\VAN}[3]{##3}\VANthebibliography}

\usepackage{graphicx}	

\usepackage{amssymb}	

\newcommand{\tar}{HS\,0218+3229}
\newcommand{\Teff}{\mbox{$T_{\mathrm{eff}}$}}
\newcommand{\logg}{\mbox{$\log g$}}
\newcommand{\HeH}{\mbox{$\log [\mathrm{He/H}]$}}
\newcommand{\cno}{\mbox{$\log\mathrm{[C/N]}$}}

\newcommand{\Ion}[2]{#1{\,\textsc{#2}}}

\newcommand{\kms}{\mbox{$\mathrm{km\,s^{-1}}$}}

\newcommand{\Rwd}{\mbox{$R_{\mathrm{WD}}$}}
\newcommand{\Mwd}{\mbox{$M_{\mathrm{WD}}$}}

\newcommand{\Msun}{\mbox{$\mathrm{M}_{\odot}$}}
\newcommand{\Rsun}{\mbox{$\mathrm{R}_{\odot}$}}

\usepackage[usenames,dvipsnames]{xcolor}

\usepackage{setspace}

\title[C/N ratio of \tar]{The C/N ratio from FUV spectroscopy as a constraint upon the past evolution of \tar}

\author[O. Toloza et al.]{
Odette~Toloza$^{1,4},$\thanks{E-mail: odette.toloza@usm.cl}
Boris T.~G{\"a}nsicke$^{2}$, 
Laura M.~Guzm{\'a}n-Rinc{\'o}n$^{3}$, 
Tom R.~Marsh$^{2}$,\newauthor 
Paula ~Szkody$^{5}$, 
Matthias R.~Schreiber$^{1,4}$, 
Domitilla de Martino$^{6}$,  
Monica Zorotovic$^{7}$,  \newauthor
Kareem El-Badry$^{8}$, 
Detlev Koester$^{9}$, and
Felipe Lagos$^{2}$\\
$^{1}$ Departamento de F\'isica, Universidad T\'ecnica Federico Santa Mar\'ia, Avenida Espa\~na 1680, Valpara\'iso, Chile\\
$^{2}$ Department of physics, University of Warwick, Gibbet Hill, Coventry CV4~7AL, UK\\
$^{3}$ Mathematics Institute, University of Warwick, Gibbet Hill, Coventry CV4~7AL, UK\\
$^{4}$ Millennium Nucleus for Planet Formation, NPF, Valpara{\'i}so, Chile\\
$^{5}$ Department of Astronomy, University of Washington, Box 351580, Seattle, WA 98195, USA\\
$^{6}$ Osservatorio Astronomico di Capodimonte, Via Moiarello 16, 80131, Napoli, Italy\\
$^{7}$ Instituto de F\'isica y Astronom\'ia, Universidad de Valpara\'iso, Av. Gran Breta\~na 1111, Valpara\'iso, Chile\\
$^{8}$ Center for Astrophysics | Harvard \& Smithsonian, 60 Garden Street, Cambridge, MA 02138, USA\\
$^{9}$ Institut f\"ur Theoretische Physik und Astrophysik, University of Kiel, 24098 Kiel, Germany
}

\date{Accepted XXX. Received YYY; in original form ZZZ}

\pubyear{2022}

\begin{document}
\label{firstpage}
\pagerange{\pageref{firstpage}--\pageref{lastpage}}
\maketitle

\begin{abstract}
Some white dwarfs accreting from non-degenerate companions show anomalous carbon and nitrogen abundances in the photospheres of their stellar components which have been postulated to be descendants of supersoft X-ray binaries. Therefore the carbon-to-nitrogen ratio can provide constraints upon their past evolution. We fit far ultraviolet spectroscopy of the cataclysmic variable \tar\ taken with the Cosmic Origins Spectrograph using Markov Chain Monte Carlo. While some parameters depend upon the amount of reddening, the carbon-to-nitrogen ratio is about one tenth of the Solar value ($\cno =-0.53^{+0.13}_{-0.14}$ and $-0.58^{+0.16}_{-0.15}$ for almost no reddening and E(B-V)=0.065, respectively, which are consistent within the uncertainties). We also provide estimates of the silicon and aluminum abundances, and upper limits for iron and oxygen. Using the measured  parameters of \tar\  we reconstruct its past using evolutionary simulations with MESA. 
We implemented Gaussian process fits to the MESA grid in order to determiner the most likely initial binary configuration of \tar. We found that an initial mass of the donor of $M_{\rm donor;i}=0.90-0.98,\Msun$ and an initial orbital period of $P_{\rm orb;i}=2.88$\,days ($P_{\rm orb;i}=3.12-3.16$\,days) for an assumed white dwarf mass of $\Mwd=0.83\,\Msun$ ($\Mwd=0.60\,\Msun$) are needed to replicate the measured parameters. These configurations imply that the system did not go through a phase of quasi-steady hydrogen-burning on the white dwarf's surface. However, it could have experienced a phase of thermal timescale mass transfer in the past if the initial mass ratio was $\geq1.5$. We predict that \tar\ will evolve into a CV with a period below the $\simeq80$\,min period minimum for normal CVs, displaying helium and hydrogen in its spectrum.
 
\end{abstract}




\section{Introduction}

Cataclysmic variables (CVs) correspond to white dwarfs stably accreting from a low mass non-degenerate companion. Mass transfer rates are too low ($10^{-10}-10^{-13}\,\Msun\,\mathrm{yr}^{-1}$) to sustain hydrogen burning on their surfaces. On the other hand, Supersoft X-ray sources, which represent one of the single-degenerate formation channels of type Ia supernovae, are thought to be white dwarfs that sustain steady nuclear fusion of hydrogen on their surfaces as a consequence of a larger mass transfer rate from a more massive companion \citep{Shara-etal-1977A&A....61..363S, van-den-Heuvel-1992A&A...262...97V, Langer-etal-2000A&A...362.1046L}. The mass-ratio ($M_{\mathrm{donor}}/\Mwd$) in these systems is roughly larger than one, which causes the mass transfer to be regulated by the thermal timescale of the donor star, leading to rates that are sufficiently high ($\gtrsim 10^{-7}\,\Msun\,\mathrm{yr}^{-1}$; \citealt{Wolf-etal-2013ApJ...777..136W}) to sustain hydrogen burning on the white dwarf surface. During this phase the white dwarf can grow in mass, which can eventually lead to a thermonuclear supernovae explosion if it approaches the Chandrasekhar mass limit. Otherwise, the Supersoft X-ray source phase will end when the mass transfer rate drops below the limit required for stable hydrogen burning (i.e. when $M_{\mathrm{donor}}/\Mwd \lesssim 1$) and the system will become a CV. However, in contrast to normal CVs that contain mostly unevolved main-sequence donors, these systems contain the stripped evolved cores of their initially much more massive companions. Henceforth, we refer to them as white dwarfs accreting from evolved main sequence companions, i.e. WD+eMS systems.

Evolutionary models indicate that WD+eMS systems descend from binaries where the white dwarf starts accreting once its donor has reached a high level of depletion of hydrogen in its core \citep[e.g.][]{Podsiadlowski-2003MNRAS.340.1214P, Kalomeni-etal-2016ApJ...833...83K}. Because of their higher initial masses, the donors in WD+eMS systems have undergone faster nuclear evolution than the donors in normal CVs. The two common mechanisms to generate energy through fusion in the centres of the stars are the proton-proton (P-P) chain and the carbon-oxygen-nitrogen (CNO) cycle. Both processes are usually present, however the stellar mass defines the physical conditions in the core, and hence which of the two processes dominates. As the mass increases, the CNO cycle becomes the main fusion reaction, fully dominating for masses $\gtrsim 1.5\,\Msun$ \citep{Wiescher-2010arXiv1011.0470W}. 

The evolutionary link between systems that, today, appear as CVs and supersoft X-ray binaries was recognised by \citet{Schenker-etal-2002MNRAS.337.1105S} for the case of AE\,Aqr. Soon thereafter, \citet{Thorstensen-etal-2002ApJ...567L..49T} identified EI\,Psc as an ultra-short period ($\simeq64$\,min) CV with an atypically hot donor star, and argued that this system started out with $M_2/M_\mathrm{wd}>1$. The current WD+eMS sample comprises several dozen systems \citep[][and references therein]{El-Badry-etal-2021MNRAS.505.2051E, El-Badry-etal-2021MNRAS.508.4106E}.

WD+eMS systems are distinguished from normal CVs since they exhibit anomalous carbon (and nitrogen) abundances in their spectra. The carbon abundance on the donor's surface can be inferred from the $^{13}$CO bands at infrared wavelengths, which is significantly lower for WD+eMS systems than in normal CVs \citep{Harrison-etal-2009AJ....137.4061H, Harrison+Marra-2017ApJ...843..152H}. As the donor's surface material is accreted onto the white dwarf the far ultraviolet spectroscopy of white dwarfs in WD+eMS systems show atypically low flux ratios of \Ion{C}{iv}/\Ion{N}{v} \citep{Gansicke-etal-2003ApJ...594..443G}. Another common property of the donors in WD+eMS systems is that they are  outliers from the mass-radius relation for main-sequence stars \citep{Knigge-2006MNRAS.373..484K}. Their radii are larger for a given mass, and therefore, Roche lobe overflow occurs at longer orbital periods, deviating from the well-established spectral type-orbital period relation of CVs. For systems with orbital periods shorter than $\simeq5-6$\,h, evolved donors are warmer than in CVs with unevolved donors, which have spectral types of M0-M5 \citep{Knigge-2006MNRAS.373..484K}. However, binary population synthesis studies predict that all CVs with orbital periods $\gtrsim 5-6$\,h are expected to have evolved secondaries \citep[e.g.][]{Beuermann-etal-1998A&A...339..518B, Podsiadlowski-2003MNRAS.340.1214P, Goliasch-etal-2015ApJ...809...80G}.

As an example, recently \citet{El-Badry-etal-2021MNRAS.505.2051E} discovered LAMOST\,J0140355+392651, an extreme member of the WD+eMS population with its donor having an effective temperature of $\Teff = 6800 \pm 100$\,K (corresponding to an F-type star) but with an orbital period of only 3.81\,h, which corresponds to an M-type donor in normal CVs.

The properties of the WD+eMS binaries suggest that some may eventually form extremely low mass (ELM) white dwarfs in detached binaries \citep{El-Badry-etal-2021MNRAS.505.2051E, El-Badry-etal-2021MNRAS.508.4106E}, or AM\,CVn systems \citep{Podsiadlowski-2003MNRAS.340.1214P}.

\tar\ was identified as a CV by \citet{Gansicke-etal-2002ASPC..261.....G} and it is located at a distance of around 500\,pc ($\varpi= 2.02\pm0.04$\,mas). Its orbital period is 7.13\,h and and it belongs to the class of WD+eMS systems. The donor is very inflated since its mass ($0.23-0.44\,\Msun$) is much lower than expected for a K-type star \citep{Rodriguez-Gil-etal-2009A&A...496..805R}. 
The optical spectrum shows emission of hydrogen and neutral helium. The amplitude of the donor’s radial velocity curve is $K_{2} = 162.4 \pm 1.4$\,km\,s$^{-1}$ \citep{Rodriguez-Gil-etal-2009A&A...496..805R}. Two outburst were detected in 1989 and 2007 and it was postulated that given the symmetry of the lightcurve, the accretion rate must be low \citep{Golyshevaetal-2013CEAB...37..345G}.

A detailed characterization of the white dwarfs in CVs is only possible from far-ultraviolet spectroscopy, since signatures of the white dwarf are diluted at optical wavelengths by the flux from the accretion disc and the donor star.
In this paper, we therefore present observations of \tar\ performed with the Cosmic Origin Spectrograph (COS) onboard \textit{Hubble Space Telescope} ({\em HST}) on 2013~May~30 aiming to derive the carbon and nitrogen abundances in their spectra, in order to confirm if it is a WD+eMS system. The average spectrum shows clear enhancement of nitrogen and depletion of carbon, confirming \tar\ as a WD+eMS. Based on Markov Chain Monte Carlo (MCMC) spectral fits we measure the carbon-to-nitrogen ratio.  This quantity combined with other measured parameters (i.e. the accretion rate, the mass and effective temperature of the donor) are used to constrain our binary simulations performed with MESA \citep{Paxton-etal-2013ApJS..208....4P, Paxton-etal-2015ApJS..220...15P, Paxton-etal-2018ApJS..234...34P, Paxton-etal-2019ApJS..243...10P}. We use Gaussian process to determine the past configuration of \tar, and determine if this system descends from a supersoft X-ray source. In addition, the MESA simulations allow us to predict the future of \tar.

\section{\textit{HST}/COS spectroscopy}
\label{sec:observations}

\begin{figure*}
    \centering
    \includegraphics[width=17cm]{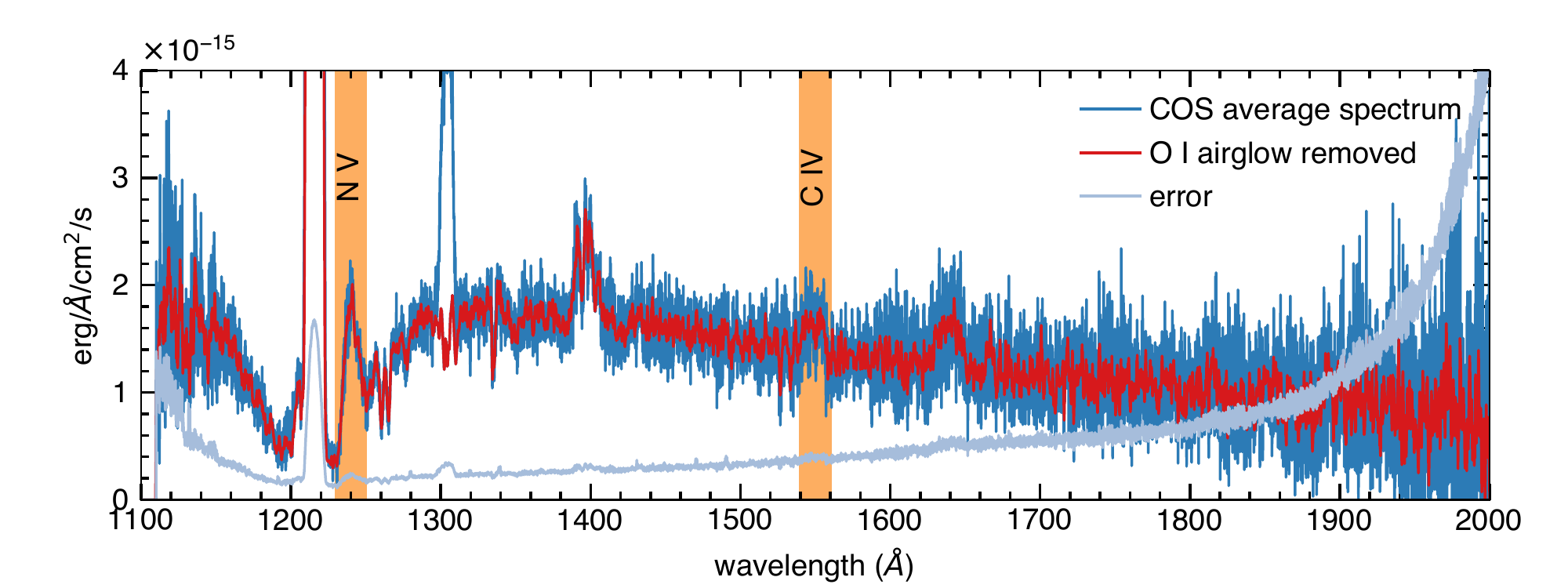}
    \caption{The COS average spectrum of \tar\ (dark blue) that shows strong emission of \Ion{N}{v} at 1240\,\AA\ and very weak emission of \Ion{C}{iv} (orange bands). These signatures are the result of past CNO-burning within the secondary star in \tar. Airglow emission of \Ion{O}{i} at 1302\,\AA\ is present in the average spectrum, however if the data taken during the daylight is removed from the average spectrum the photospheric absorption of \Ion{O}{i} is exposed (red spectrum; convolved with a 10-point boxcar). The flux error is shown in gray, illustrating the rapid deterioration in the signal-to-noise ratio at wavelengths $>1850$\,\AA. \label{fig:spectrum}}
    
\end{figure*}

\tar\ was observed as part of of the programme 12870. A total of 119.3\,min of far-ultraviolet time-tagged spectroscopy of \tar\ were obtained on 2012~December~22 over three spacecraft orbits resulting in five separate exposures, which cover less than a full orbital cycle, i.e. 3.65\,h.
The G140L grating centred at 1105\,\AA\ was used for the observations, providing a wavelength coverage of $1150-1800$\,\AA\ with roughly $0.75$\,\AA\ resolution. Table~\ref{tab:obs} summarises the observations and the time-averaged spectrum is shown in Figure~\ref{fig:spectrum}. 

\begin{table}  
\begin{center}
\caption[Observations]{Log of the {\em HST}/COS observations of \tar\ taken in time-tag mode and using the G140L/1105 grating. \label{tab:obs}}
\begin{tabular}{c c c c c}
\hline
\hline
 ID & orbit & date	& UT Time 	& exp. time (s)	\\
\hline
lc1v19o4q	&1&2012-12-22	&06:31:07	&1956\\
lc1v19ocq	&2&2012-12-22	&07:49:21	&1753\\
lc1v19ohq	&2&2012-12-22	&08:20:19	&853\\
lc1v19ojq	&3&2012-12-22	&09:25:03	&856\\
lc1v19olq	&3&2012-12-22	&09:41:14	&1740\\
\hline
\end{tabular}
\end{center}
\end{table}

The average COS spectrum is dominated by the emission of the white dwarf. However, it shows an additional flat and featureless flux component that is clearly evident in the core of the broad Ly\,$\alpha$ at 1216\,\AA. This flux component has been seen in the far-ultraviolet spectroscopy of the many other CVs \citep[e.g.][]{Gansicke-etat-2005ApJ...629..451G,Toloza-etal-2016MNRAS.459.3929T}, however the contribution of this second component is small ($\lesssim$ 20 per cent). Its origin is unknown, some suggestions are that the flux could originate from the hot innermost region of the accretion disc, or a boundary/spreading layer on the white dwarf \citep{Godon+Sion-2005MNRAS.361..809G, Godon-etal-2008ApJ...679.1447G}.

The spectrum shows signatures of CNO processing, i.e.  a strong emission line of \Ion{N}{v} at 1240\,\AA, while the emission line of \Ion{C}{iv} at 1550\,\AA\ is much weaker than normally seen in CVs. In addition, the spectrum shows emission lines of \Ion{He}{ii} at 1640\,\AA\ and \Ion{Si}{iv} at 1400\,\AA\  \citep{Mauche-etal-1997ApJ...477..832M}.
The other emission lines correspond to Earth's airglow emission lines of Ly\,$\alpha$ at 1216\,\AA\ and \Ion{O}{i} at 1302\,\AA, the latter disappears if the data are taken during \textit{HST}'s night time. The \textsc{time-tag} mode records the position and the arrival time of the photons, and thus allows a correction for the \Ion{O}{i} airglow contamination. We used the {\sc timefilter} task from the \textsc{costools} package version 1.2.3 to exclude the photons recorded during daylight. The  extraction and calibration\footnote{We used the reference files associated with the observations which can be downloaded directly from \href{ftp://ftp.stsci.edu/cdbs/lref}{ftp://ftp.stsci.edu/cdbs/lref}} of these night side spectra were performed with the \textsc{x1dcorr} task from the \textsc{costools} package. An average spectrum which excludes the airglow of \Ion{O}{i} is combined by using tasks from \textsc{calcos} pipeline version 3.3.9 (see Figure~\ref{fig:spectrum}). To preserve the maximum signal-to-noise ratio of the COS observations, we only substituted the night side data in the region affected by O\,{\sc i} airglow in the average spectrum. With the removal of the airglow of \Ion{O}{i}, the spectrum reveals the white dwarf's photospheric absorption line of \Ion{O}{i}. The COS spectrum also shows absorption lines of \Ion{C}{ii} at 1335\,\AA, \Ion{Al}{ii} at 1640\,\AA, and \Ion{Si}{ii} at 1260,1265\,\AA, and at 1527,1533\,\AA.

\subsection{Radial velocity and rotational broadening of the white dwarf}
\label{sec:broadening}

\begin{figure}
    \centering
    \includegraphics[width=8cm]{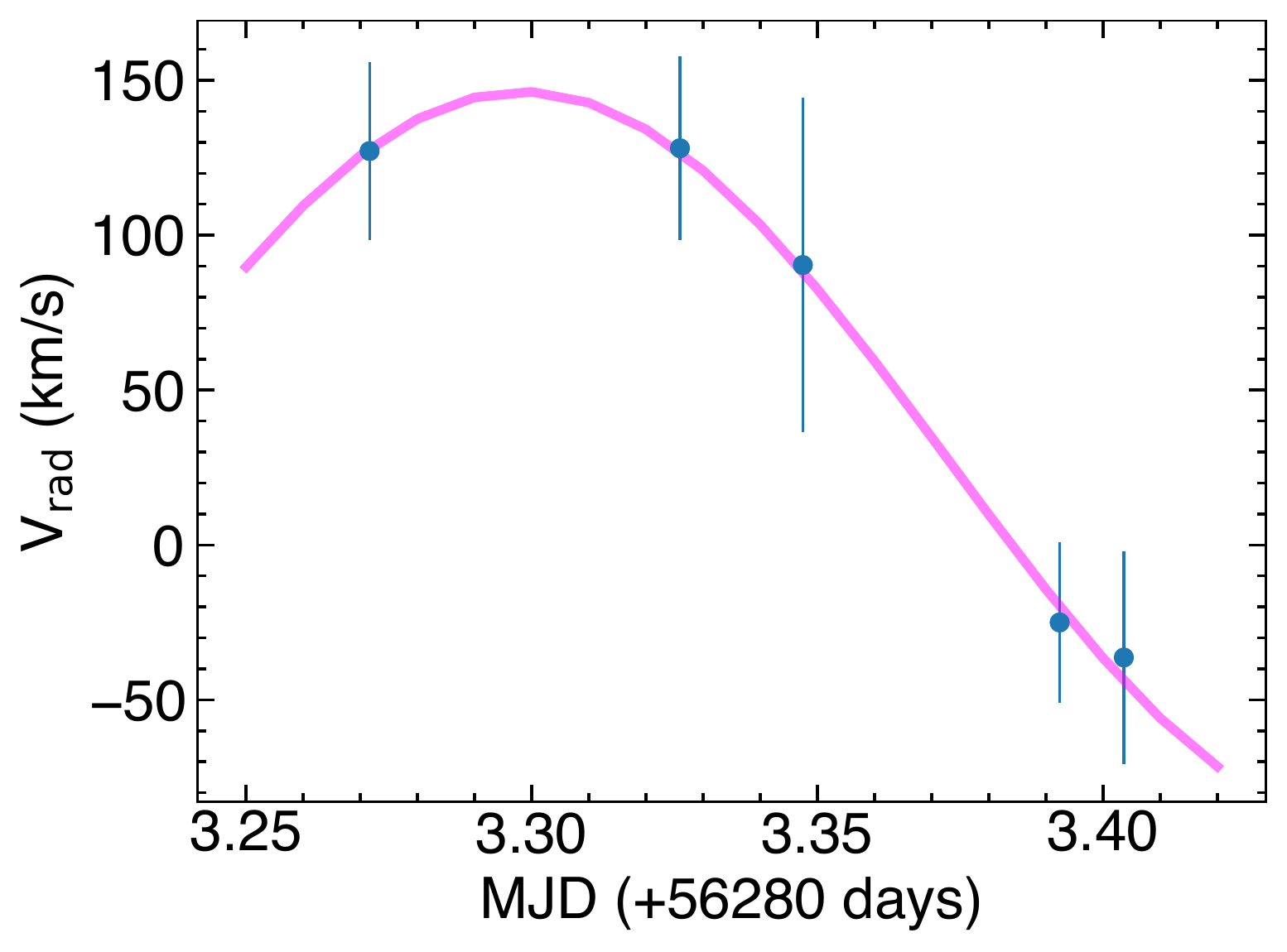}
    \caption{The radial velocity measurements (blue dots) obtained from the \Ion{Si}{ii} line at 1265\,\AA\  for the five sub-exposures listed in Table~\ref{tab:obs} are shown together with the best fit sinusoidal (pink) function with amplitude of 118.3~km\,s$^{-1}$ (see text for details). \label{fig:vrad}}
\end{figure}

\citet{Rodriguez-Gil-etal-2009A&A...496..805R} estimated the radial velocity of the white dwarf to be $100.3\pm1.2$~km\,s$^{-1}$, thus the orbital motion might have affected the observed line profiles since the coverage of the COS data ($\simeq3$~hours) is almost half an orbital cycle. 
We attempted to measure the radial velocity of the white dwarf in \tar\ using the five individual COS exposures ($\lesssim$ 30\,min; Table \ref{tab:obs}).

Each of the individual exposures was fitted using the Levenberg-Marquardt algorithm with best fit model obtained from the MCMC fits to the average spectrum is used. This best fit model was corrected by the instrumental broadening around the \Ion{Si}{ii} 1265\,\AA\ line ($R=2226$). Each \textsc{x1d} exposure provides a radial velocity and $v~\mathrm{sin}i$. For \textsc{lc1v19ojq} the fit did not converge, thus, excluding this point, the weighted-mean and weighted-standard deviation for the rotation corresponds to $v~\mathrm{sin}i= 179 \pm 27$~km\,s$^{-1}$. The results of the radial velocities are shown in Figure~\ref{fig:vrad}. We fit a sinusoidal function to these measurements to estimate the radial velocity amplitude ($K_{\mathrm{WD}}$), the systemic velocity ($\gamma_{\mathrm{sys}}$), and the ephemeris \cite[the orbital period was fixed to the value shown in Table~\ref{tab:parameters};][]{Rodriguez-Gil-etal-2009A&A...496..805R}. Results are $K_{\mathrm{WD}}= 118 \pm 31$\,km\,s$^{-1}$ and $\gamma_{\mathrm{sys}}=28.07 \pm 24.98$\,km\,s$^{-1}$. Our estimate of the radial velocity amplitude of the white dwarf agrees with the estimate of \citet[][ $K_{\mathrm{WD}}=100.3\pm1.2$\,km\,s$^{-1}$]{Rodriguez-Gil-etal-2009A&A...496..805R}  within the uncertainties.

\section{Spectral Fits}
\label{sec:fits}
We aim to determine the metal abundances seen in the COS spectrum of \tar. To accomplish this, we fitted the average spectrum using the Markov Chain Monte Carlo (MCMC) method \citep{Foreman-Mackey-etal-2013PASP..125..306F} with synthetic models for white dwarfs using the atmospheric code of \citet{Koester-2010MmSAI..81..921K}. 

The atmosphere of the white dwarf was modelled with four main parameters: the effective temperature (\Teff), the surface gravity (\logg), the helium-to-hydrogen ratio (\HeH), and the metal abundances ($Z$). The Eddington flux ($H_{\rm Edd}$) of the synthetic models were scaled to the observed spectrum as,

\begin{multline}
    F_{\rm obs} = 4\,\pi\,\left(\varpi \times \Rwd\right)^{2}\,\times\,H_{\rm Edd} \times  [10^{-0.4 \times (\mathrm{ext}+R(V))\times E(B-V)}] + f_{\rm k},
    \label{eq:model}    
\end{multline}

where $\varpi$ is the parallax, \Rwd\ is the white dwarf radius, the term within the square brackets reddens the synthetic models by $E(B-V)$ using an extinction model where ($R(V)$ is the ratio of total to selective extinction; $R(V) = A(V) / E(B - V)$), and finally $f_{\mathrm{k}}$ is a extra constant flux. The parameters in this model (equation \ref{eq:model}) are explained in detail below together with the constraints and priors that are imposed on them:

\begin{itemize}

    \item parallax ($\varpi$): $\varpi$ can be constrained using the \textit{Gaia} parallax. While the Early Data Release~3 (EDR3; $\varpi=2.02 \pm 0.04$\,mas) provides a more precise measurement than the Data Release~2 (DR2; $\varpi=1.93 \pm 0.06$\,mas), we still define a very conservative parallax prior. Using the inverse parallaxes, \tar\ located at a distance of $D=485-535$\,pc. Given that the fractional distance accuracy decreases with increasing distance, we also derived the distance using the Bayesian inference method from \citet{Bailer-Jones-etal-2021AJ....161..147B}, which gives a geometric and geophotometric distances that are consistent with the inverse parallax within uncertainities, and thus we estimate the distance as the inverse parallax. Therefore, during the fits, we set a flat prior on the parallax allowing the best fit value to be within $1.87-2.06$\,mas. 
    
    \item white dwarf radius (\Rwd): the \Rwd\ is a function of \logg\ and \Teff\ through the mass-radius relation for white dwarfs. We obtained the mass-radius relation for white dwarfs with hydrogen-rich atmospheres by interpolating the cooling models from \citet{Bedard-etal-2020ApJ...901...93B} with thick hydrogen layers of $M_{\rm H}/\Mwd=10^{-4}$, which are available from the University of Montreal website\footnote{\href{http://www.astro.umontreal.ca/~bergeron/CoolingModels}{http://www.astro.umontreal.ca/$\sim$bergeron/CoolingModels}}.

    \item effective temperature (\Teff): The \textit{HST}/COS spectrum does not show the broad quasimolecular ($H_{2}$ and $H^{+}_{2}$) absorption features around 1400\,\AA\ and 1600\,\AA\ that result from radiative collisions of excited atomic hydrogen ($H^{+}$) and unexcited neutral hydrogen atoms ($H$), respectively \citep{Koester-etal-1985A&A...142L...5K}. These features are clearly seen in hydrogen-rich white dwarfs with temperatures $\lesssim$15\,000\,K, and they are absent if the atmosphere is helium rich \citep{Hoskin-etal-2020MNRAS.499..171H}. Therefore, the absence of these features suggests either the atmosphere is helium rich or that the effective temperature of the white dwarf should be hotter than 15\,000\,K. A constraint on the upper limit for the white dwarf's effective temperature is given by the ionization of the photospheric metal lines. We clearly see \Ion{Si}{ii} in the COS spectrum, but not \Ion{Si}{iii}, which starts to be visible for temperatures above 20\,000\,K. Therefore, we constrained
    the effective temperature between $15\,000-20\,000$\,K. 
    
    \item surface gravity (\logg): The \textit{Gaia} parallax places a strong constraint on the surface gravity. In addition, a rough estimate of the surface gravity can be obtained from the dynamical white dwarf mass estimate \citep[\Mwd=$0.44-0.65$\,\Msun;][]{Rodriguez-Gil-etal-2009A&A...496..805R} using the mass-radius relation (see above). With the same mass-radius relation explained above, we find that $\Mwd=0.44-0.65\,\Msun$ results in a surface gravity of \logg=$7.6-8.1$\,dex. However, we allowed a more relaxed range to exploit the accuracy of the \textit{Gaia} parallaxes. Therefore,  we set a flat prior on the surface gravity of $7.0-8.5$ during the fits. 
    
    \item helium-to-hydrogen ratio (\HeH): The optical spectrum of \tar\ shows emission lines of the Balmer series and of \Ion{He}{I} (see Figure~2 in \citealt{Rodriguez-Gil-etal-2009A&A...496..805R}) from the disc. Therefore, the material accreted onto the atmosphere of the white dwarf is replenished with hydrogen and helium. The abundant presence of helium in the white dwarf atmosphere contributes to the shape of the ``pseudo-continuum'', being more noticeable for wavelengths shorter than 1150\,\AA. Therefore, we adopted a mixed hydrogen/helium atmosphere allowing the less dominant element (either hydrogen or helium) to be as low as $-5.0$\,dex\ with respect to the more abundant one. For example, if the white dwarf has a hydrogen dominated atmosphere, the limit on helium will be $\HeH=-5.0$\,dex.
    
    \item metals ($Z$):  We include the elements that show the strongest absorption lines in the COS spectrum that could affect the opacity in the white dwarf's atmosphere. The COS spectrum shows clear absorption lines of silicon (\Ion{Si}{ii} at 1260, 1265\,\AA; 1527,1533\,\AA), and carbon (\Ion{C}{ii} at 1335\,\AA). It also shows a weak line of aluminum (\Ion{Al}{ii} at 1670\,\AA), as well as one line of oxygen (\Ion{O}{i} at 1302\,\AA, which is revealed once the airglow emission lines of oxygen have been removed (Figure~\ref{fig:spectrum}). Moreover, the spectrum contains two broad features around $1125-1135$\,\AA\ which could be the combination of the many iron lines in that region. We include nitrogen since the emission of \Ion{N}{v} at 1240\,\AA\ (Figure~\ref{fig:spectrum}) indicates that the accreting material is nitrogen-enriched (see section \ref{sec:nitrogen}). The strongest nitrogen absorption features are expected around 1135\,\AA\ and 1493\,\AA. In summary, carbon, nitrogen, oxygen, aluminum, silicon, and iron were included, with which we constrained the logarithm of the element relative to a base element ($XY$; either hydrogen or helium) to be always negative ($\mathrm{log}(Z/XY)<0$).
    
    \item reddening, $E(B-V)$:  The far-ultraviolet wavelengths are largely affected by extinction, which increases with distance. The 3D dust extinction maps, i.e. the STructuring by Inversion the Local Interstellar Medium (\textit{Stilism}\footnote{\href{https://stilism.obspm.fr/}{https://stilism.obspm.fr/}} and the updated version using \textit{Gaia}/2MASS\footnote{\href{https://astro.acri-st.fr/gaia\_dev/}{https://astro.acri-st.fr/gaia\_dev/}}), show that \tar\ is considerably affected by reddening at its distance. To compensate for reddening, the emitted flux from the star needs to be larger than if there were no reddening, which can be achieved by either a hotter white dwarf (increase of the effective temperature) and/or a white dwarf with a larger surface area (where larger radius decreases the surface gravity and the white dwarf mass). However, the exact amount of reddening is still subject to uncertainty given the accuracy of the extinction maps and it is unclear how reddening would affect the metal abundances. Therefore, we perform two separate MCMC fits: (1) we fit reddening imposing a flat prior $E(B-V)=0.0-0.085$\,mag taken from Stilism, and (2) we fix reddening to a maximum value of 0.065\,mag,  taken as a reference value from from \citep[$E(B-V)=0.0612 \pm 0.0010$;][]{Schlafly+Finkbeiner-2011ApJ...737..103S} and $E(B-V)=0.98\times E(g-r)$=0.0686 \citep{Green-etal-2019ApJ...887...93G}.
    
    \item the second component ($f_{\mathrm{k}}$): In past studies, this has been modelled with either a blackbody, a power law or with a constant flux  \citep[e.g.][]{Gansicke-etat-2005ApJ...629..451G, Szkody-etal-2010ApJ...710...64S, Pala-etal-2017MNRAS.466.2855P}. However, when the flux is mainly dominated by the white dwarf, the three models do not significantly affect the estimates of the other parameters when fitting the the short ultraviolet range covered by COS spectroscopy of white dwarfs in CVs. Here, we adopt a flux component constant in flux, as it reduces the total number of free parameters, and we set that it has to be positive ($f_{\mathrm{k}}>0$).
    
\end{itemize}

The mixing length parameter (ML2/$\alpha$) in 1-dimensional convection models depends on the composition of the atmosphere. For pure-hydrogen-atmosphere white dwarfs the values are lower \citep[e.g. ML2/$\alpha=0.6$; ][]{Bergeron-etal-1995ApJ...449..258B} than for pure-helium-atmosphere white dwarfs \citep[e.g. ML2/$\alpha=1.25$;][]{Bergeron-etal-2011ApJ...737...28B, Voss-etal-2007A&A...470.1079V}. Due to the strong opacity of hydrogen, a helium-dominated atmosphere with some hydrogen will appear as a hydrogen-dominated atmosphere \citep{Koester-etal-2005A&A...439..317K}. However, the dominant atmospheric composition of the white dwarf in \tar\ is unknown. Here, we treat it as a mixed-atmosphere, and set the parameter to ML2/$\alpha=1.0$ \citep{Cukanovaite-etal-2019MNRAS.490.1010C}. Besides hydrogen and helium, we included 626 of the strongest metal lines of carbon, nitrogen, oxygen, aluminum, silicon and iron and their opacities ($1000-15\,000$\,\AA) during the computation of atmosphere structure.

Finally, we applied the COS/G140L instrumental broadening to the models. The resolving power of the G140L grating is $1500-4000$, increasing linearly with wavelength. In order to speed up the fitting proccess, we convolved the synthetic models with a Gaussian kernel with a resolving power of 2700 (calculated around the carbon and nitrogen lines, i.e. around 1500\,\AA). 
The identified emission lines (described in Section~\ref{sec:observations}) were masked out in the fits. 

We use the \textsc{emcee} ensemble sampler \citep{Foreman-Mackey-etal-2013PASP..125..306F}. We set 80 and 40 walkers to sample the parameter space for the run where the reddening is free and fixed, respectively. For each iteration of MCMC the white dwarf synthetic models are computed in order to calculate the likelihood, where the likelihood function is defined as $-\chi^{2}/2$.

\subsection{Autocorrelation analysis of the chains}

Sampling the posterior distribution over the parameter space is computationally expensive. In every MCMC iteration, 80 (40 for the MCMC fit with fixed reddening) synthetic models are computed in parallel. On average, the time to compute an atmospheric model running the code from \citet{Koester-2010MmSAI..81..921K} takes $\simeq5-10$\,min and, therefore, each MCMC iteration takes  $\simeq5-10$\,min. As a consequence, it becomes unfeasible to build chains with steps of several thousands. Since the MCMC potentially requires a large number of independent samples, it is necessary to estimate in advance the number of steps needed to ensure convergence. The integrated autocorrelation time $\tau$ quantifies the effective number of independent samples from a chain \citep{Foreman-Mackey-etal-2013PASP..125..306F}, and therefore, it provides the number of evaluations required to obtain independent samples.

\begin{figure}
    \centering
    \includegraphics[width=0.8\columnwidth]{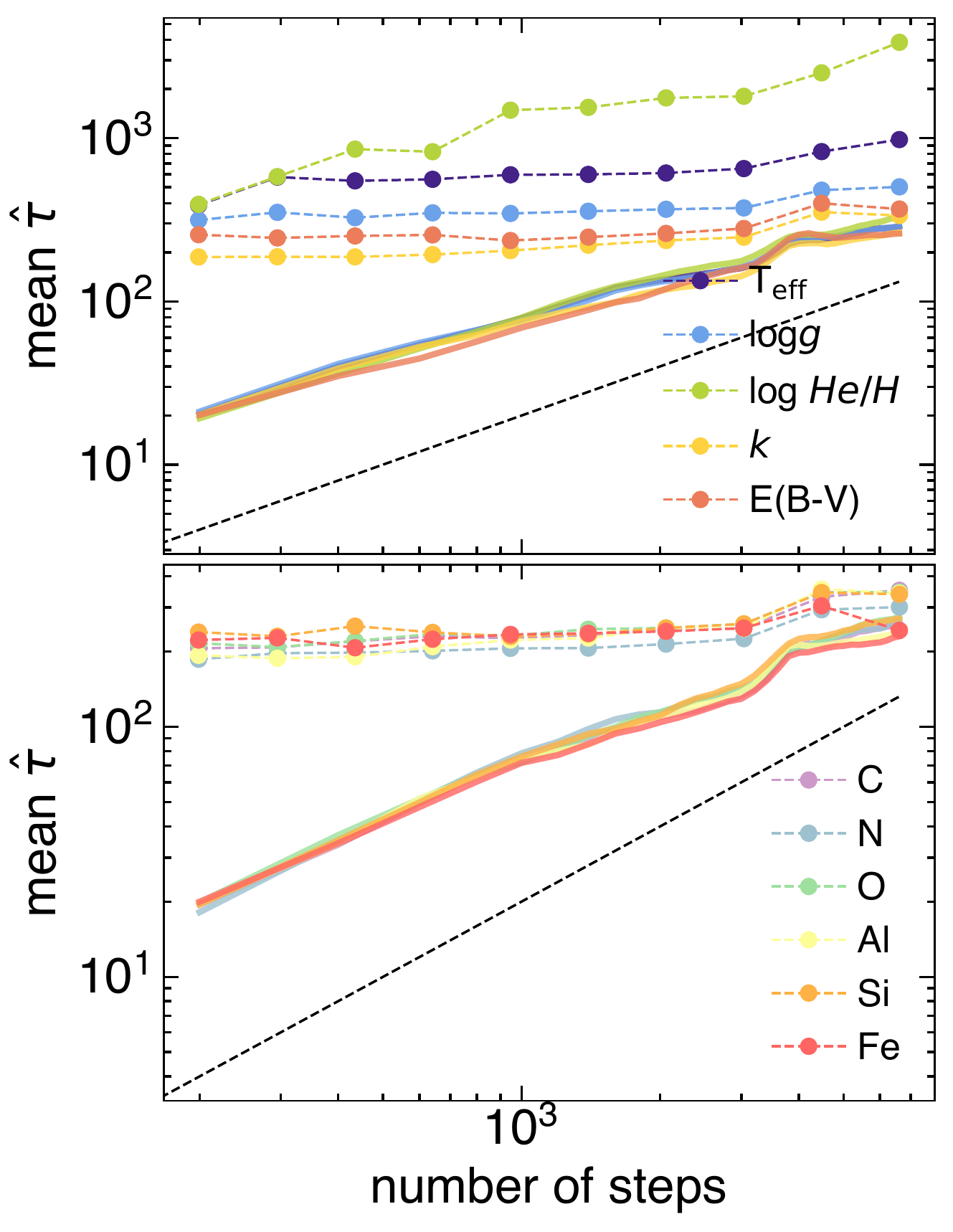}
    \caption{Autocorrelation time estimated at different lengths of the chain. The solid lines correspond to the estimator $\hat{\tau}_{\rm ST}$, calculated at each 100th step. The black dashed lines correspond to the autocorrelation time $\hat{\tau}_{\rm GP}$ of the fitted Gaussian process. The dashed lines (number of steps/50) represents the lower limit for which the length of the chain would provide an reliable estimate of $\tau$. In contrast to $\hat{\tau}_{\rm GP}$, the convergence of $\hat{\tau}_{\rm ST}$ is very slow. \textit{Top:} The effective temperature (\Teff), the surface gravity (\logg), the second component ($f_\mathrm{k}$), and the reddening ($E(B-V)$) have well-constrained $\hat{\tau}_{\rm GP}$ with values less than 1000 steps. In contrast, the helium-to-hydrogen ratio (\HeH) is the only parameter that does not present a clear value for $\hat{\tau}_{\rm GP}$, which aligns with the fact that this parameter can not be strongly constrained from the observations. \textit{Bottom:} The abundances require only $\hat{\tau}_{\rm GP}\simeq200$~steps providing a large number of independent samples within the MCMC run. }
    \label{fig:tau}
\end{figure}

\begin{figure}
    \centering
    \includegraphics[width=0.8\columnwidth]{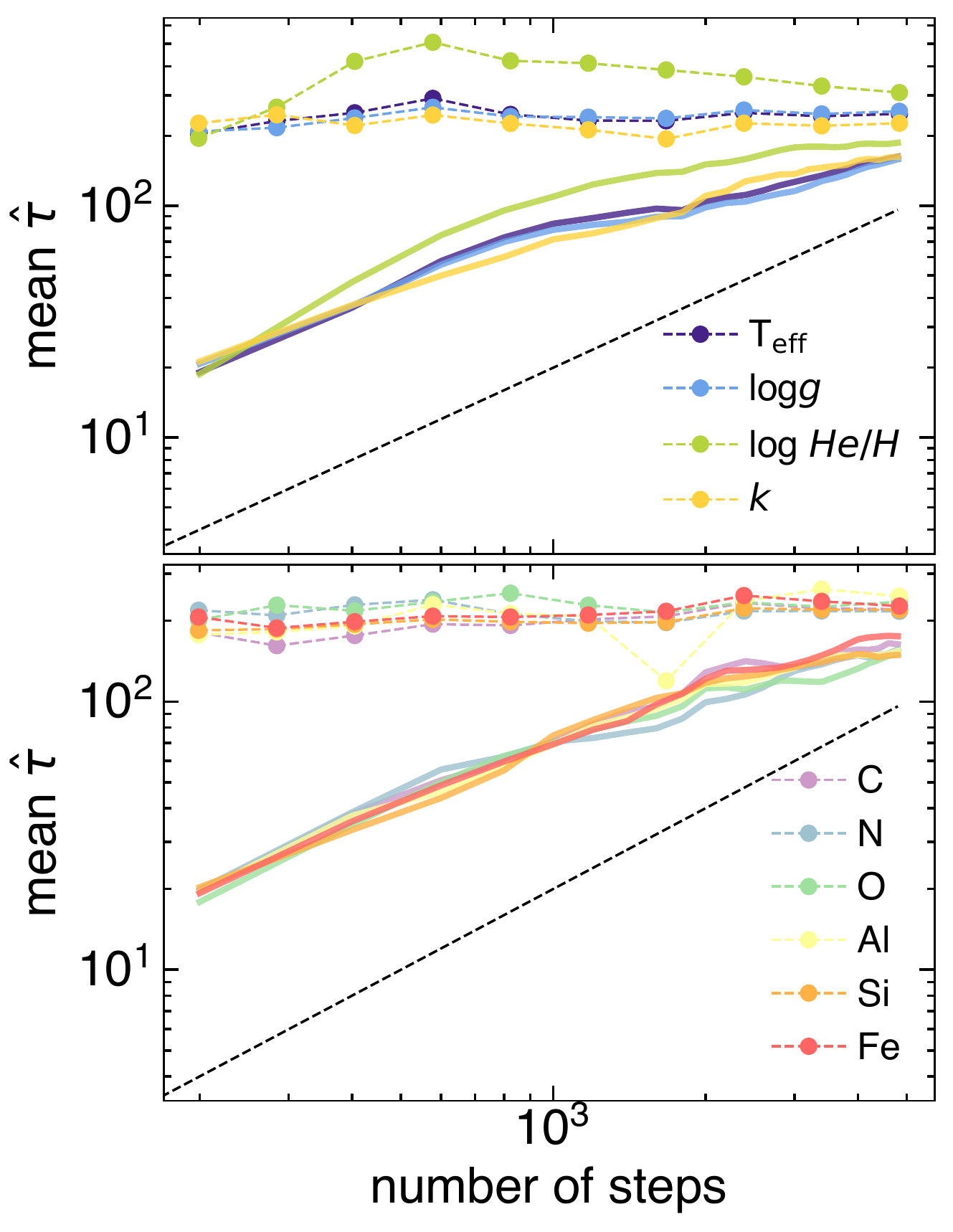}
    \caption{Same as Figure~\ref{fig:tau} but with $E(B-V)$ being fixed, which, results in the effective temperature and surface gravity requiring a lower $\hat{\tau}_{\rm GP}$.}
    \label{fig:tau_ebv}
\end{figure}

An accurate estimate of the autocorrelation time also requires long chains. However, due to the computational limitations, we derived an estimate of $\tau$ and evaluated its convergence to the true value using two methods. First, we used an estimator proposed by \citet{Goodman+Weare-2010-MCMC}, labelled as $\hat{\tau}_{\rm ST}$, and calculated it at every 100th step. However, the convergence of this estimator was slow. Second, we fitted the chains using a Gaussian Process (GP) with a mixture of exponential functions as the kernel. The GP was computed using the celerite model, a fast algorithm that scales linearly with the size of the data set \citep{Foreman-Mackey-etal-2017AJ....154..220F}. The autocorrelation time for the GP was computed based on the parameters of the kernel and labelled as $\hat{\tau}_{\rm GP}$. Both methods were applied to each parameter of the model, and the estimated $\tau$ was averaged for the 80 (40) chains. Figures \ref{fig:tau} and \ref{fig:tau_ebv} show the speed of convergence of both estimators of the autocorrelation time for the parameters, which are represented with solid and dashed lines for $\hat{\tau}_{\rm ST}$, and $\hat{\tau}_{\rm GP}$, respectively. It is clear that $\hat{\tau}_{\rm ST}$ shows a very slow convergence and requires additional steps to provide a reliable estimate of $\tau$. Therefore, we use $\hat{\tau}_{\rm GP}$ for the analysis of the convergence. In general, it is clear that all the chains require more than 100 steps for the walkers to erase the correlation from its starting position, and some of the parameter converge more quickly than others. The roughly constant $\tau$ around 300 indicates that the constant flux, $f_{\mathrm{k}}$ is the parameter that converges the quickest. Similarly, the abundances (bottom panels) converge very quickly requiring $\hat{\tau}_{\rm GP}\simeq300-400$ steps. In contrast, the helium-to-hydrogen ratio (\HeH) is the parameter with the largest $\hat{\tau}_{\rm GP}$, and with the length of the MCMC chains that we could computationally afford, it is not possible to report convergence on this parameter. The $\hat{\tau}_{\rm GP}$ of the effective temperature and surface gravity are larger for the case when reddening is included as an additional free parameter (Fig.~\ref{fig:tau}), while $\hat{\tau}_{\rm GP}$ is roughly 200 steps when reddening is fixed (Fig.~\ref{fig:tau_ebv}).

As the stretch move in the Markov chain satisfies the detailed balance condition \citep{Goodman+Weare-2010-MCMC}, we can consider the 80 (40 for the case of the MCMC run excluding reddening) chains as independent. Therefore, the length of the MCMC run of a few (5-10) times the autocorrelation time for each of the parameters will provide a sufficient number of independent samples in the marginalized distributions. Thus, with the exception of \HeH, all the parameters have converged  and provide their corresponding $\hat{\tau}_{\rm GP}$ in less than 1000 steps ensuring enough samples for the analysis of the parameters.

\subsection{Results}
\label{sec:results_fits}
\begin{figure*}
    \centering
    \includegraphics[width=17cm]{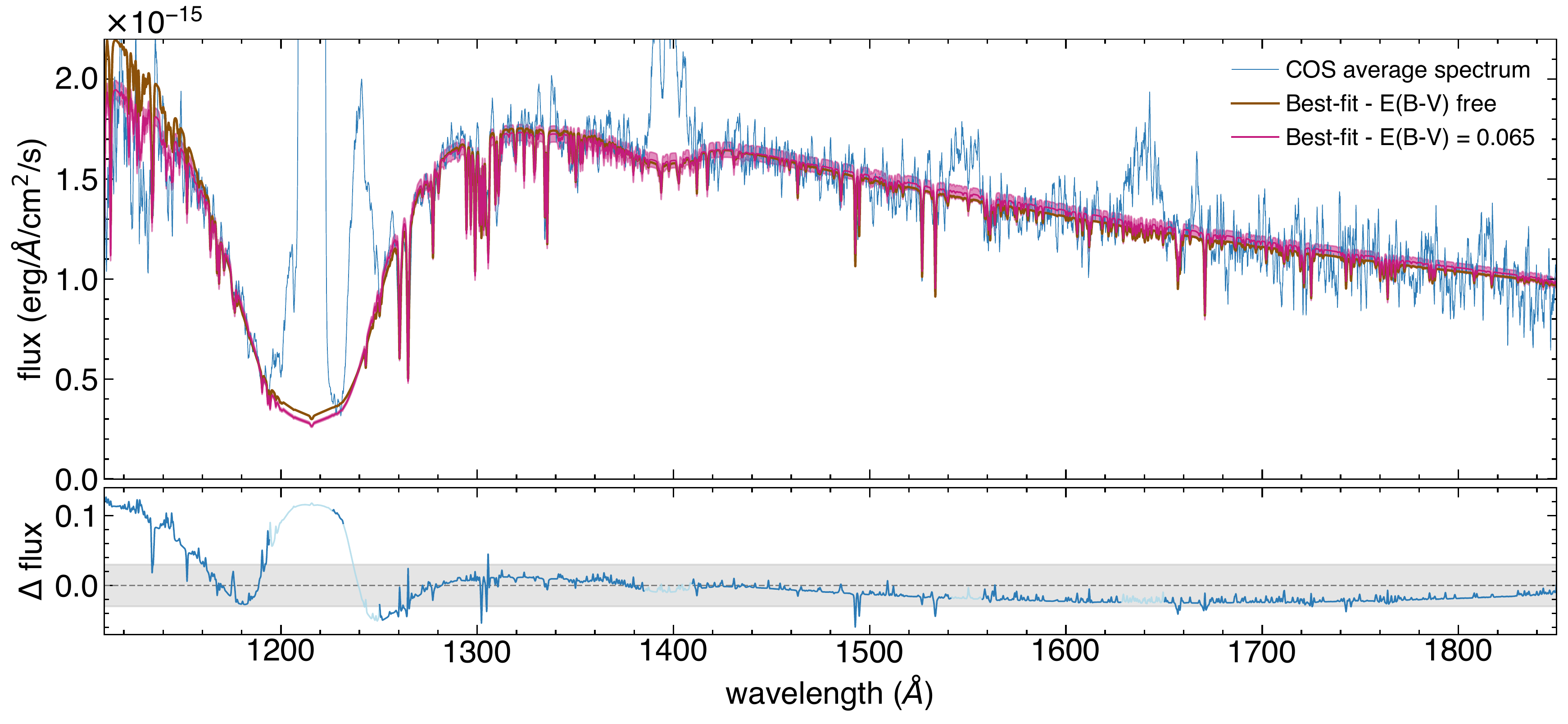}
    \caption{\textit{Top:} The best MCMC-fits for free reddening (brown) and reddening fixed to $E(B-V)=0.065$\,mags (magenta) are overplotted on the COS average spectrum of \tar\ (blue). The best-fit model is computed from the mean of the last 80 and 40 samples of the MCMC runs, respectively. The fixed reddening MCMC fit also shows the 99~per cent confidence interval using the normal-distribution (pink band) of the samples of the last step. \textit{Bottom:} Fractional difference between the models in blue, where the light blue regions represents the data that was masked out during the fits. Grey band shows three per cent fractional difference.}
    \label{fig:fit}
\end{figure*}

The top panel in Figure~\ref{fig:fit} shows the best fits for MCMC fits where reddening is free (brown), and fixed to 0.065\,mag (magenta). The best-fit spectrum corresponds to the median of the last samples of the respective MCMC runs. To illustrate the variance of the MCMC fit, the models of the walkers at the last step are shown within the pink band. The bottom panel in Figure~\ref{fig:fit} shows the fractional difference between these two best fit models. With the exception of the core of the strongest absorption lines, for wavelengths longer than $\sim$1270\,\AA\ the two best-fit models differ by less than three per cent (grey band). In contrast, at shorter wavelengths the fractional difference rises to $\simeq10$~per cent in two problematic regions. The former corresponds to wavelengths shorter than 1160\,\AA\  which could be affected by additional broadening that Ly\,$\alpha$ experiences by perturbations from neutral helium \citep{Gansicke-etal-2018MNRAS.481.4323G}, which is not included in our synthetic models. However, the amount of neutral helium decreases towards higher effective temperatures and so for effective temperatures hotter that 16\,000\,K, these broadening effects are minimal. The second problematic region corresponds to the core of the broad Ly\,$\alpha$ absorption profile, which is contaminated by the geocoronal airglow Ly\,$\alpha$, which was masked out, and several data points in that region were removed during the fitting process.

The best-value parameters (Table \ref{tab:parameters}) are considered as the median (50$^{\mathrm{th}}$), and the errors are the 16$^{\mathrm{th}}$ and 84$^{\mathrm{th}}$ percentiles of the 1-dimensional marginalized distributions (Figures \ref{fig:corner} and \ref{fig:corner_ebv} for reddening free and fixed, respectively). As the errors are intrinsic to the MCMC method, they are purely statistical in nature. 

\begin{table}
\centering
\caption{Best MCMC fit parameters (first block), where the abundances are in number of atoms relative to hydrogen and in brackets are shown the values relative to solar  \citep{Asplund-etal-2009ARA&A..47..481A}. The second block lists the white dwarf mass and radius as well as the accretion rate and the carbon-to-nitrogen ratio, which are derived from the best fit parameters. Finally, the third block lists the parameters of \tar\ determined by \citet{Rodriguez-Gil-etal-2009A&A...496..805R}. The effective temperature of the donor is estimated from its spectral type (see section~\ref{sec:past_evolution}). Parameters marked with an asterisk (*) provide constraints to evolutionary simulations (see section~\ref{sec:mesa})\label{tab:parameters}}.
\begin{spacing}{1.5}
    \begin{tabular}{l c c}
    \hline
    \hline
Parameter & $E(B-V)$ Free & $E(B-V)=0.065$  \\
\hline
$\varpi$ [mass] & $1.99_{-0.060}^{+0.058}$ & $ 2.04_{-0.023}^{+0.011}$\\
$T_{\mathrm{eff}}$ [K] & $19271_{-193}^{+145}$ & $18740_{-168}^{+126}$ \\
$\log g$ [cgs] & $8.36_{-0.07}^{+0.06}$ &  $7.95_{-0.04}^{+0.03}$\\
$\log [\mathrm{He/H}]$ & $0.01_{-0.12}^{+0.10}$ -- [1.07]& $-0.14_{-0.18}^{+0.14}$ -- [1.07]\\
$f_{\mathrm{k}}$\, [$\times$10$^{-16}$ erg\,s$^{-1}$ \AA$^{-1}$\,cm$^{-2}$] & $2.94_{-0.13}^{+0.12}$ & $2.59_{-0.13}^{+0.14}$\\
$E(B-V)$ [mag] & $0.005_{-0.004}^{+0.007}$ &  -- \\
$\log [\mathrm{C/H}]$ & $-4.70_{-0.16}^{+0.12}$ -- [-1.13] & $-4.72_{-0.11}^{+0.09}$ -- [-1.15]\\
$\log [\mathrm{N/H}]$ & $-4.15_{-0.02}^{+0.34}$ -- [0.02] & $-4.14_{-0.13}^{+0.13}$ -- [0.03]\\
$\log [\mathrm{O/H}]$ & $-4.29_{-0.11}^{+0.01}$ -- [-0.98] & $-4.19_{-0.21}^{+0.18}$ -- [-0.88]\\
$\log [\mathrm{Al/H}]$ & $-6.37_{-0.14}^{+0.26}$ -- [-0.82] & $-6.23_{-0.35}^{+0.29}$ -- [-0.68]\\
$\log [\mathrm{Si/H}]$ & $-5.15_{-0.06}^{+0.28}$ -- [-0.66] &  $-5.04_{-0.06}^{+0.08}$ -- [-0.55]\\
$\log [\mathrm{Fe/H}]$ & $-5.48_{-0.10}^{+0.21}$ -- [-0.98] & $-5.11_{-0.20}^{+0.17}$  -- [-0.61]\\
\hline
$M_{\mathrm{WD}}\,[\Msun]$ & $0.84_{-0.05}^{+0.04}$& $0.60_{-0.02}^{+0.02}$ \\
$R_{\mathrm{WD}}\,[\Rsun]$ & $0.0101_{-0.0005}^{+0.0006}$ & $0.0135_{-0.0003}^{+0.0003}$\\
$\log \dot{M}\,[\Msun\,\mathrm{yr}^{-1}]^*$ & $-9.66^{+0.04}_{-0.04}$ & $-9.40^{+0.02}_{-0.01}$\\
$\cno^*$ &$-0.53^{+0.13}_{-0.14}$ -- [-0.07]&$-0.58^{+0.16}_{-0.15}$-- [-0.02]\\
\hline
\multicolumn{1}{l}{} &\multicolumn{2}{c}{literature value}\\
\hline
SpT donor & \multicolumn{2}{c}{K5}\\
$T_{\mathrm{eff, donor}}$ [K]$^*$ & \multicolumn{2}{c}{$4382\pm 326$}\\
$M_{\mathrm{donor}}$ [\Msun]$^*$ & \multicolumn{2}{c}{$0.23-0.44$}\\
$P_{\mathrm{orb}}$ [h] & \multicolumn{2}{c}{$7.13351186\pm 0.00000002$}\\
\hline

    \end{tabular}
\end{spacing}
\end{table}

\begin{figure*}
    \centering
    \includegraphics[width=2\columnwidth]{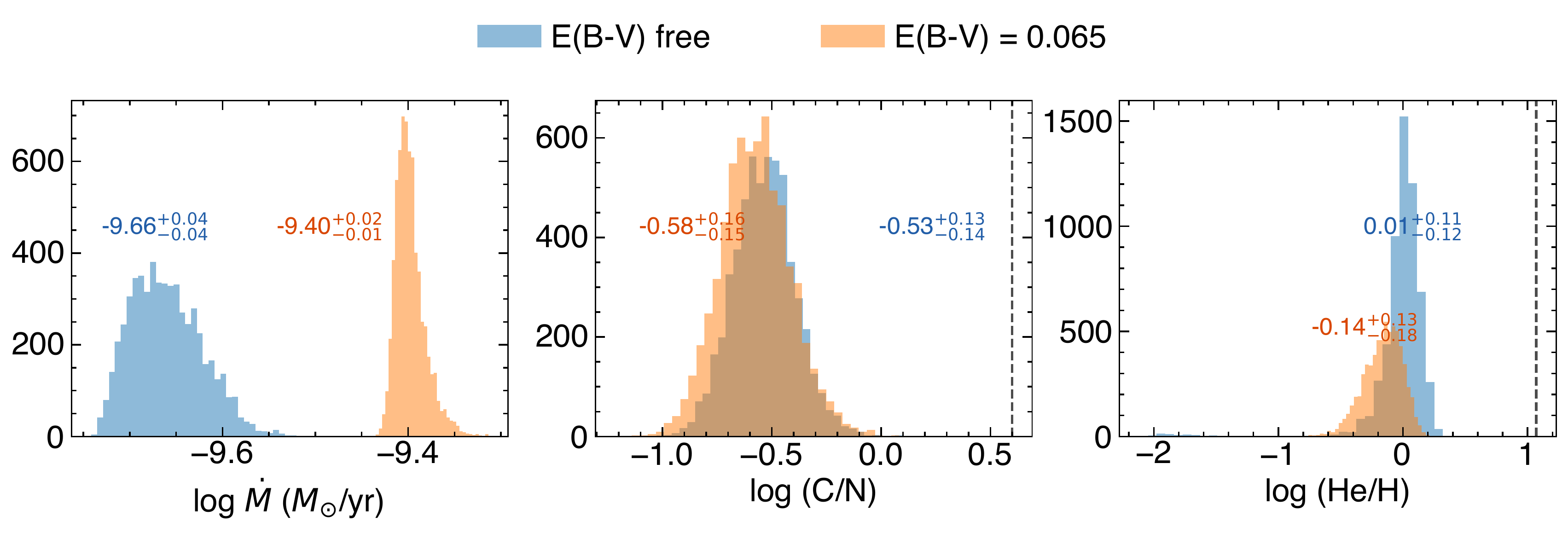}
    \caption{Accretion rate ($\log\dot{M}$), carbon-to-nitrogen ratio ($\log\mathrm{[C/N]}$), and helium-to-hydrogen ratio ($\log\mathrm{[He/H]}$) obtained from the MCMC samples for the cases when reddening is free (blue) and fixed (orange). The accretion rate strongly depends on the samples of effective temperature and surface gravity which both are significantly affected by reddening (see text for details). These parameters provide constraints to the binary simulations (see Section \ref{sec:mesa}). The dashed line represents the solar value.}
    \label{fig:observables}
\end{figure*}

While the parallax has been tightly constrained to be within $1.87-2.06$\,mas, the samples cluster towards larger values. The distributions of the effective temperature and surface gravity follow pseudo-Normal distributions, and they are highly correlated (Pearson correlation coefficient larger than 0.85 for both MCMC fits). 
This correlation arises because the broad absorption line of Ly\,$\alpha$ is very sensitive to changes in effective temperature and surface gravity and hence is a key feature to determine these parameters from far-ultraviolet spectroscopy. High temperatures are required to sustain ionization of hydrogen, in contrast, high surface gravities lead to high atmospheric pressure which decreases the degree of ionization of hydrogen. As a consequence, the balance between effective temperature and surface gravity causes a strongly correlated degeneracy. The mass and radius distributions of the white dwarf can be determined using the samples of \Teff\ and \logg\ from MCMC along with the above mass-radius relation. The mean values and uncertainties of these distributions are shown in Table~\ref{tab:parameters}. For the case where reddening was fixed, the mass estimate is within the uncertainties of the dynamical mass estimate from  \citet[\Mwd=0.44--0.65\,\Msun;][]{Rodriguez-Gil-etal-2009A&A...496..805R}. A secondary parameter extracted from MCMC fits is the accretion rate. Based on the compressional heating produced on the white dwarf due to the accreted material onto its surface, we can estimate the time-averaged secular accretion rate  (equation~1 from \citealt{Townsley+Gansicke2009ApJ...693.1007T}). This analytic expression is a function of the mass and effective temperature of the white dwarf. Similarly as for the mass and radius of the white dwarf, we use the \Teff\ and \logg\ samples to build the accretion rate samples (left panel in Figure~\ref{fig:observables}). Because higher reddening affects the surface gravity and effective temperature, the secular accretion rate is also higher. From these samples the 15th and 84th percentiles dictates their uncertainties (Table~\ref{tab:parameters}).

The second component, $f_\mathrm{k}$, converged very quickly and its value is very well constrained. The distribution of the samples of reddening (Figure~\ref{fig:corner}) clusters towards low value. The median of this distribution corresponds to 0.005\,mag which is much lower than the estimate from the 3-dimensional dust maps from stilism.

The \HeH\ does not show a clear correlation with any parameter, except for the effective temperature, which shows a non-marginal degree of correlation. The Pearson correlation coefficient are 0.72 and 0.74 for the cases when reddening is free and fixed, respectively. The best values for helium-to-hydrogen ratios suggest a balanced mixture of hydrogen and helium. The samples are shown in the right panel of Figure~\ref{fig:observables}, where the effect of a higher reddening is to push the samples towards slightly lower values. However, in both cases the helium abundance is lower than the solar value (dashed line in Fig~\ref{fig:observables}). We stress that this method does not provide a reliable estimate of \HeH\ as it instead does the spectroscopic fits of the Balmer and helium lines seen in the optical range in single white dwarfs with mixed atmospheres. The spectrum is very sensitive if the helium-to-hydrogen ratios tend to helium dominated atmospheres, in contrast, the hydrogen dominated atmospheric models do not differ. Therefore, we report the helium abundances derived from these ratios as upper limits.

An accurate estimate of the chemical abundances relies on a precise estimate of \HeH\, since the presence of helium makes the atmosphere more transparent and hence, for a given amount of a chemical element, its absorption lines are stronger in helium-dominated atmospheres than for hydrogen-dominated atmospheres. However, the ratio between the elements (middle panel of Figure~\ref{fig:observables}) is a quantity that is not affected by the inaccurate determination of \HeH. An accurate determination of \HeH\ is beyond the scope of this work.

Finally, the chemical elements do not show clear correlation among them (Figures \ref{fig:corner} and \ref{fig:corner_ebv}). In general, all of their samples follow a Normal distribution, and their medians are larger for the CNO elements. The comparison of the abundances between the two MCMC fits shows that the best values are not strongly affected by reddening.

\subsection{Chemical abundances and the CNO elements}
\label{sec:metals}

\begin{figure*}
    \centering
    \includegraphics[width=2.1\columnwidth]{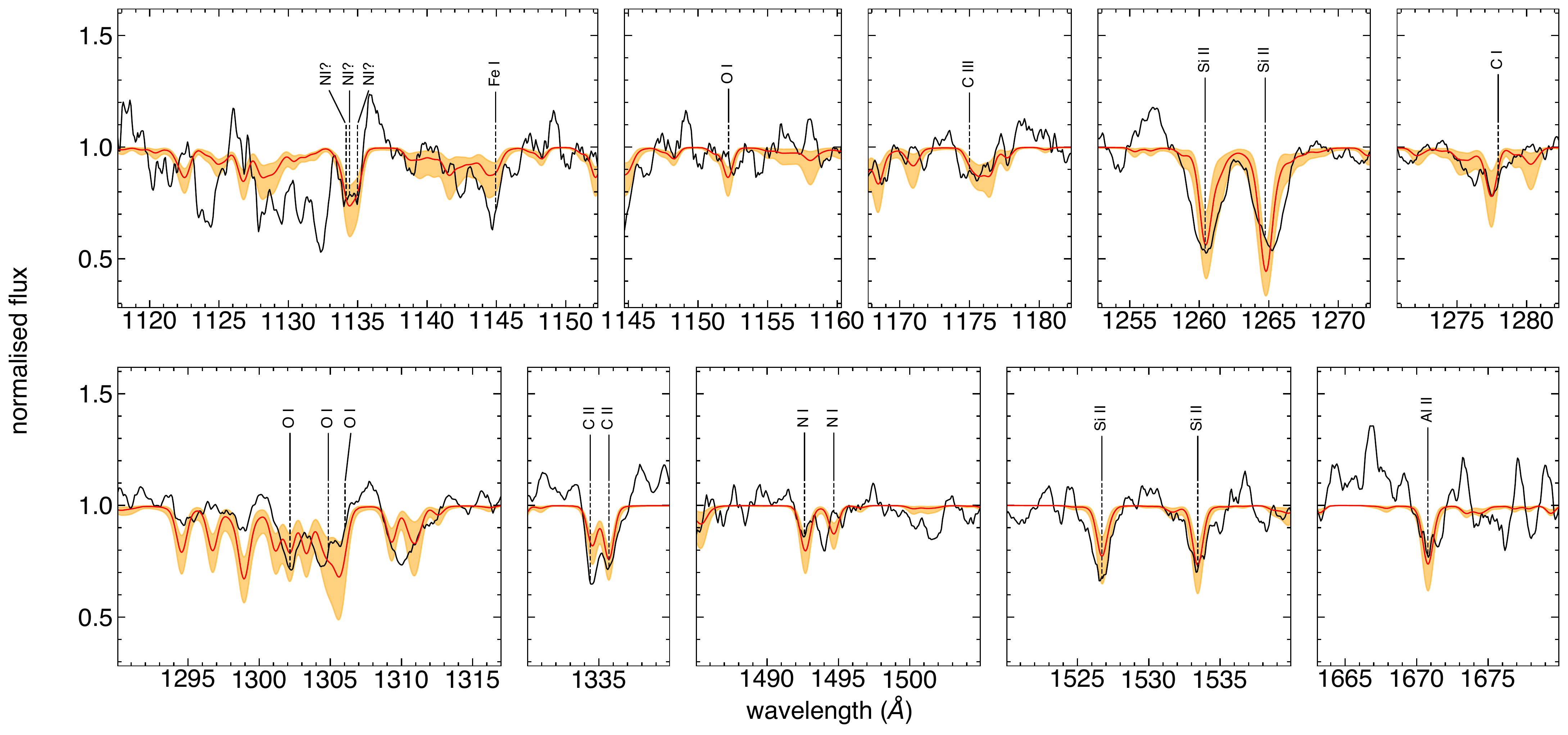}
    \caption{Regions where the transitions of carbon, nitrogen, and oxygen are found in the far-ultraviolet wavelength regime. For completeness, the strongest absorption lines of silicon and aluminum are shown. Overplotted in red is the best fit model computed using the median of the best fit parameters (Table~\ref{tab:parameters}). The orange represents a $\pm$0.5 
    \,dex change in the abundances that illustrate the sensitivity that the abundance values has on the line depth. Note that the error in the abundance estimates are $<0.5$\,dex. The spectrum has been convolved with a 10-point boxcar for display purposes}
    \label{fig:panels}
\end{figure*}

The atoms of different abundances provide additional opacities affecting the white dwarf atmosphere. Studies from high-resolution spectroscopy of FGK-type stars have gathered thousands of stars from which their abundances are measured. The most common abundances in the literature \citep[][and references therein]{Jofre-etal-2019ARA&A..57..571J} range from light and $\alpha$-elements (C,  O,  Na,  Mg,  Al,  Si,  Ca,  Ti), to the iron-peak elements (Sc, V, Cr, Mn, Fe, Co, Ni). Some of these elements do not show transitions in the ultraviolet (e.g Na and Ca) or the strongest transitions are very weak (e.g \Ion{Mg}{ii} at 1734.85, 1737.62, 1750.65, 1757.18\,\AA; or several lines of \Ion{Ti}{iii} around 1290\,\AA). Thus, to account for opacities we considered only those chemical abundances that are clearly seen in the \textit{HST}/COS spectrum (i.e. Al, Si and Fe). It is worth to mention, that strong absorption of \Ion{Fe}{ii} and \Ion{Al}{ii} lines could rise from accretion gas that veils the white dwarf. This so-called iron curtain is seen in systems with high inclination like OY\,Car, SDSS\,J100658.41+233724.4, and DV\,UMa \citep{Horne-etal-1994ApJ...426..294H, Pala-etal-2017MNRAS.466.2855P}. Thus, as \tar\ has an inclination of $i = 59^{\circ}\pm 3^{\circ}$ \citep{Rodriguez-Gil-etal-2009A&A...496..805R} we can not rule out the presence of a low density column causing additional absorption. An accurate determination of any of these abundances apart from the CNO products is beyond the scope of this paper. However, the marginalised distributions of Al, Si, and Fe are shown in Figures~\ref{fig:corner} and \ref{fig:corner_ebv}, which show the median values of these samples are lower that $\simeq-5.0$ with aluminum being the lowest abundance. It is worth noting that the Si abundance has smaller uncertainties as expected given that the spectrum has several strong absorption lines (Table~\ref{tab:parameters}).  All these abundances give sub-solar values with the exception of nitrogen which is close to solar (values in brackets in Table~\ref{tab:parameters}).

The carbon, nitrogen and oxygen in the CNO cycle act as catalysts to convert hydrogen into helium. While the reactions in the cold CNO cycles are primarily radiative proton capture, the slowest reaction is $^{14}$N(p,$\gamma$)$^{15}$O. As a result the majority of the carbon in the core will be transformed into nitrogen \citep{Wiescher-etal-2010}. The samples of the carbon-to-nitrogen ratio from the MCMC fits are shown in Figure~\ref{fig:observables}, corroborating that nitrogen in the white dwarf atmosphere is enhanced in comparison to carbon.

In general, the abundances do not show any correlation between them (the Pearson correlation coefficients are lower than 0.3). The strong absorption lines are shown in Figure~\ref{fig:panels}. Here, we explain in more detail as absorption lines from CNO products.

\subsubsection{Carbon}
\label{sec:carbon}
The strongest line of carbon is the doublet of \Ion{C}{ii} at 1335\.\AA. However, the \Ion{C}{ii}~1335.7\,\AA\ transition has larger statistical weight than \Ion{C}{ii}~1334.5\AA\, and thus the former should be stronger in the white dwarf atmosphere. However, the COS spectrum contradicts the above statement, indicating that there is a contribution from interstellar (IS) absorption. Another transition free of IS contamination that is seen in hot white dwarfs in CVs \citep[e.g. RR  Pic;][]{Sion-etal-2017AJ....153..109S} is the absorption of \Ion{C}{iii} at 1175\,\AA. However, the spectrum of \tar\ shows a very weak absorption line (Figure~\ref{fig:panels}) due to the low degree of ionization. Based on these lines, the MCMC fits gives measurements of ${\rm{log}}~[C/H]= -4.70^{+0.12}_{-0.16}$, and ${\rm{log}}~[C/H]=-4.72^{+0.09}_{-0.11}$ when reddening is considered as a free and fixed parameter, respectively. (Table~\ref{tab:parameters}).

\subsubsection{Nitrogen}
\label{sec:nitrogen}

Using the atomic data from VALD\footnote{\href{http://vald.astro.uu.se/}{http://vald.astro.uu.se/}} we found that the strongest nitrogen lines that become visible for the temperature range of \tar\ are in the triplet of \Ion{Ni}{i} 1134.16,1134.41,1134.98\,\AA\ (Figure~\ref{fig:panels}). These lines are also present in the interstellar medium and fall within the region of the Fe lines and thus complicating the fits. The next stronger lines are the doublet 1492.62,1494.68\,\AA\ lines of \Ion{Ni}{i} (Figure~\ref{fig:panels}) which contribute to an unidentified feature at 1494\,\AA. With the absence of clear strong features, rough estimates are provided from the MCMC fits  (i.e. log~$[N/H] = -4.15^{+0.34}_{-0.02}$, and log~$[N/H] = -4.14^{+0.13}_{-0.13}$ for a free and fixed reddening, respectively; Table~\ref{tab:parameters}), which should be considered as upper limits.

\subsubsection{Oxygen}
\label{sec:oxygen}
The \Ion{O}{i} line at 1152.2\,\AA\ is diluted within the noise of the spectrum, and therefore the variance of the spectroscopic data place an upper limit on the strength of this line (Figure~\ref{fig:panels}). 
With the geocoronal emission line removal, several absorption lines of \Ion{O}{i} around 1302\,\AA\ become visible in the spectrum (Figure~\ref{fig:panels}). The transition of \Ion{O}{i} at 1302.2\,\AA\ has been identified in the interstellar medium, and so, the \Ion{O}{i} at 1304.9\,\AA\ turns into the most reliable line to determine the oxygen abundance. The MCMC fits provide estimates of log~$[O/H] = -4.29^{+0.01}_{-0.11}$, and log~$[O/H]=-4.19^{+0.18}_{-0.21}$ for free and fixed reddenning, respectively (Table~\ref{tab:parameters}).

\medskip

In what follows we aim to determine the initial configurations of the post common envelope binaries that can reproduce the above described measurements and the ones from the literature.  In particular, we explore if \tar\ has experienced thermal timescale mass transfer when it was a young CV. To that end we performed binary star simulations with MESA that are described in the next section.

\section{The evolutionary history}

\label{sec:mesa}

Investigating possible evolutionary pathways that led to the formation of \tar\ and its siblings requires detailed simulations that are best performed with MESA. Before we describe the set-up and results of the simulations we performed, we provide a brief review of the current understanding of close white dwarf binary formation and evolution. 

\subsection{Stability of mass transfer and angular momentum loss}

The stability of mass transfer depends on the response of the donor to mass loss which in turn depends on the structure of the envelope (i.e. convective envelopes tend to expand, while radiative envelopes shrink rapidly). Given that the dynamical timescale ($\tau_{\mathrm{dyn}}$) is shorter that the Kelvin-Helmholtz timescale ($\tau_{\mathrm{KH}}$), the first response of the donor to mass loss is to change its radius adiabatically to reestablish hydrostatic equilibrium. 
Thermal equilibrium might then be established on the much longer thermal time scale.

The stability of mass transfer is quantified by comparing the mass-radius exponents of the star for both time scales discussed above ($\zeta_{\mathrm{ad}}$ and $\zeta_{\mathrm{th}}$) with the Roche lobe, i.e. $\zeta$=d ln\,R/ d ln\,M and $\zeta_{\mathrm{RL}}$=d\,ln\,R$_{\mathrm{RL}}$/ d ln\,M \citep[e.g.][]{Soberman+vandenHeuvel-1997A&A...327..620S}.  
The response of the Roche-lobe radius to mass transfer depends on the mass ratio and how much mass is lost in the process (i.e. how conservative the mass transfer is). Therefore, one can define a critical mass ratio for a given donor mass and assuming either conservative or non-conservative mass transfer.
In case mass transfer is not conservative, one further has to take into account the angular momentum loss of the binary due to the mass loss. Because this angular momentum loss only occurs in the case of mass transfer it is called consequential angular momentum loss (in contrast to systemic angular momentum loss). 

\begin{figure}
    \centering
    \includegraphics[width=1\columnwidth]{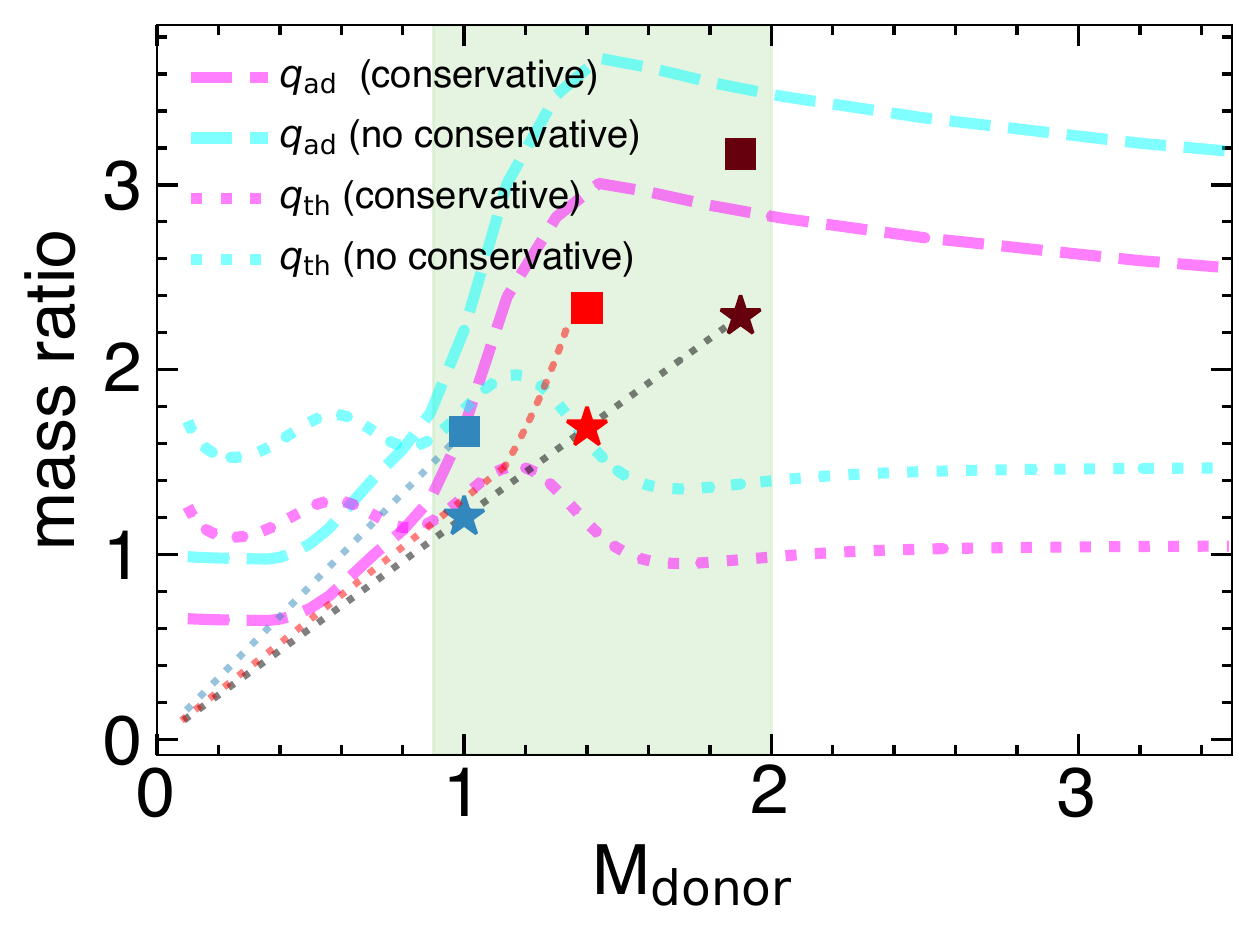}
    \caption{Mass ratio stability of dynamical mass transfer (dashed cyan and pink lines) and thermal mass transfer (dotted cyan and pink lines) assuming either conservative (pink) or non-conservative (cyan) mass transfer. The evolution of MESA tracks are shown for a white dwarf mass of \Mwd=0.83\,\Msun\ (stars) and \Mwd=0.6\,\Msun\ (squares), where their initial orbital periods correspond to 2.7~days and 3.0~days, respectively. The initial masses of the donor star in these tracks are 1.0\,\Msun\ (blue symbols), 1.4\,\Msun\ (red symbols), and 1.9\,\Msun\ (brown symbols). The green shaded region represents the donor masses that span the MESA grid (see section~\ref{sec:setup}).}
    \label{fig:Qcrit}
\end{figure}

We used the adiabatic exponents for the zero age main sequence from \citet{Ge-etal-2015ApJ...812...40G} to derive the critical mass ratio for dynamical stability ($q_{\mathrm{ad}}$).
The critical mass ratio for thermal stability ($q_{\mathrm{th}}$) was computed using the analytic expression of the mass-radius relation for the zero age main sequence from \citet{Tout-1996MNRAS.281..257T}. These resulting stability limits are shown in Fig.~\ref{fig:Qcrit} for both adiabatic (dashed) and thermal (dotted) responses of the donor. The color indicates whether the mass transfer was assumed to be conservative (magenta) or non-conservative (cyan). In case of the latter, we assumed the angular momentum of the expelled material to be equal to the specific angular momentum of the white dwarf as derived by \citet{King+Kolb-1995ApJ...439..330K}.

Figure~\ref{fig:Qcrit} allows to distinguish three cases of mass transfer. Firstly, if the mass transfer is dynamically stable and the donor is able to recover its thermal equilibrium ($\zeta_{\mathrm{th}} < \zeta_{\mathrm{RL}}$ and $\zeta_{\mathrm{ad}} < \zeta_{\mathrm{RL}}$), accretion is driven by systemic angular momentum loss and secularly stable. In this case, hydrogen accumulates on the white dwarf surface until it rapidly burns in a thermonuclear runaway process (known as a nova eruption), i.e. the mass tranfer is not conservative. Second, if the mass transfer is dynamically stable but the donor is not able to recover its thermal equilibrium ($\zeta_{\mathrm{th}}<\zeta_{\mathrm{RL}}< \zeta_{\mathrm{ad}}$), the mass transfer rate is determined by the thermal readjustments of the donor. This leads to much larger mass transfer rates than in the first case. If the accretion rate exceeds a certain limit ($\dot{M}_{\mathrm{HB}}$) the accreted hydrogen can stably burn on the white dwarf surface causing the binary to become a supersoft X-ray source \citep{Paczynsli+Zytkow-1978ApJ...222..604P,Iben-1982ApJ...259..244I, Fujimoto-1982ApJ...257..767F, Livio-1989ApJ...341..299L, Cassisi-etal-1998ApJ...496..376C, Shen-etal-2007ApJ...660.1444S, Nomoto-etal-2007ApJ...663.1269N}. In this case, the mass transfer is highly conservative. Towards higher accretion rates above the $\dot{M}_{\mathrm{HB}}$ limit, the amount of matter that is burnt into helium is limited by a maximum luminosity, defining a upper bound that can be approximated by $3\times\,\dot{M}_{\mathrm{HB}}$ \citep{Wolf-etal-2013ApJ...777..136W}. Lastly, if the donor is unable to readjust its hydrostatic equilibrium within the Roche lobe on a dynamical timescale ($\zeta_{\mathrm{ad}} > \zeta_{\mathrm{RL}}$), the mass transfer is dynamically unstable. In this case, mass transfer proceeds on the dynamical time scale, which is much shorter than the thermal time scale of the white dwarf and therefore, instead of accumulating on the surface of the white dwarf, the accreted gas forms an envelope around the white dwarf at the base of which hydrogen burning may be triggered. In other words, the white dwarf converts into a red-giant \citep{Paczynsli+Zytkow-1978ApJ...222..604P, Nomoto-etal-1979PASJ...31..287N}.

If mass transfer is thermally and dynamically stable, the mass transfer can be driven either 
by nuclear evolution or by (systemic) angular momentum loss. 
Most cataclysmic variables are stable against dynamical and thermal mass transfer 
and mass transfer is driven by angular momentum loss. The two main mechanisms of angular momentum loss are gravitational wave radiation \citep{Paczynski-1967AcA....17..287P} and magnetic braking generated by the donor star \citep{Verbunt+Zwaan-1981A&A...100L...7V, Rappaport-etal-1983ApJ...275..713R, Mestel+Spruit-1987MNRAS.226...57M, Kawaler-1988ApJ...333..236K, Andronov-2003ApJ...582..358A}. It is thought that magnetic braking becomes inefficient (or even vanishes completely) when the systems reach orbital periods of $\simeq3$\,h, which is thought be a consequence of the donor becoming fully convective. As a direct consequence, there is an abrupt decrease in the mass transfer rate, which allows the donor to relax to its equilibrium radius causing the systems to detach, crossing the well-known period gap \citep[$P_\mathrm{gap} \simeq 2.15-3.18$\,h;][]{Knigge-2006MNRAS.373..484K}. 
The subsequent evolution is driven by angular momentum loss through the emission of gravitational waves and mass transfer is resumed when the donor fills its Roche lobe again at the lower edge of the period gap (i.e. $P_{\mathrm{orb}}\simeq2.15$\,h). 
When the donor mass becomes too low to sustain hydrogen burning, it becomes a hydrogen-rich degenerate
object, i.e. a brown dwarf. From this point on, the donor expands in response to the mass loss, leading to
an increase in the orbital period. The so-called orbital period minimum is then a natural consequence of the
transition from a fully convective main sequence star to a brown dwarf. The minimum period has been measured to be  \citep[$P_{\mathrm{min}}=76.1\pm1$\,min;][]{Knigge-2006MNRAS.373..484K}.

With the aim of constraining the evolutionary history of \tar\, we performed binary evolution simulations for initially massive secondary stars that evolve into cataclysmic variables using MESA version 12778. The exact set-up of the simulations is described in detail in the next section.

\subsection{Evolutionary simulations with MESA}
\label{sec:setup}

We adopted the standard model of the evolution of cataclysmic variables. We have set the exponent of the magnetic braking prescription \citep{Rappaport-etal-1983ApJ...275..713R} to $\gamma=3$. Magnetic braking is operating a long as the star has a radiative core, and it is not turned off at high effective temperatures of the donor. The mass transfer rate through Roche lobe overflow was described with the prescription of \citet{Ritter-1988A&A...202...93R}. The micro- and macro-physics of the donor star account for the evolution of low mass stars (i.e. M$\lesssim$2M$_{\odot}$). We included the nuclear reaction networks of \textsc{pp\_and\_cno\_extras} to account for the hot CNO cycle. The solar abundances from \citet{Grevesse+Sauval-1998SSRv...85..161G} provide the OPAL opacities (\textsc{kappa\_file\_prefix=gs98}). We used the Type2 opacity tables, where the base metallicity for the opacity tables was set to \textsc{Zbase=0.02}. In the mixing-length theory of convection of \citet{Cox+Giuli-1968pss..book.....C} we set the mixing length parameter to \textsc{mixing\_length\_alpha=2}. However, for stars with masses in the range of $1.0-2.0$\,M$_{\odot}$, the energy produced by the CNO cycle is higher than with the PP chain, and thus convection is driven in the center of the star. As the star evolves, the convective core expands slightly, causing a discontinuity of the hydrogen profile at the edge of the convective core. A natural consequence is the development of a semiconvective region above the convective core due to opacities contributing to the radiative temperature gradients \citep{Paxton-etal-2018ApJS..234...34P, Paxton-etal-2019ApJS..243...10P}. Thus, we used the Ledoux criterion (\textsc{use\_Ledoux\_criterion}=True), and included semiconvection with the highest efficiency \textsc{alpha\_semiconvection=1.0}, to approach to the Schwarzchild criterion \citep{Langer-1991A&A...252..669L}. 

We defined the stopping condition to be when the donor becomes a brown dwarf 
(\textsc{star\_mass\_min\_limit}=0.08\,\Msun; \citealt{Dieterich-etal-2014AJ....147...94D}).

We assumed fully non-conservative mass transfer, i.e. all the material accreted on the white dwarf is expelled from the system due to nova eruptions (i.e. $\Mwd={\rm constant}$; $\textsc{mass\_transfer\_beta} = 1.0$ and \textsc{do\_jdot\_ml}=True), except when systems reach the hydrogen-burning stability described by \citet{Wolf-etal-2013ApJ...777..136W}. This mass transfer threshold for hydrogen-burning ($\dot{M}_{\mathrm{HB}}$) is a function of the mass of the white dwarf, i.e. the more massive the white dwarf, the higher the accretion rate needed to sustain a shell stably burning hydrogen on the white dwarf surface \citep{Nomoto-etal-2007ApJ...663.1269N, Shen-etal-2007ApJ...660.1444S, Wolf-etal-2013ApJ...777..136W}. If it occurs, this phase is usually short, and we denote it as pseudo-conservative mass transfer scenario, where a fraction of mass is retained by the accretor (we here assumed that 90 per cent of the transferred material is retained by the white dwarf, i.e. $\textsc{mass\_transfer\_beta} = 0.1$, which prevents numerical problems \citealt{Parsons-etal-2015MNRAS.452.1754P}).

In summary, for systems going trough a hydrogen burning phase, the evolution implemented in MESA is divided into three legs: at the beginning of mass transfer the accretion rate is well below the limit for hydrogen burning and we therefore assumed fully non-conservative mass transfer. Later, mass transfer might possibly reach the $\dot{M}_{\mathrm{HB}}$ threshold, at which point the system enters a pseudo-conservative mass transfer phase until the mass transfer either surpasses the $3\times\,\dot{M}_{\mathrm{HB}}$ threshold or falls below the $\dot{M}_{\mathrm{HB}}$ limit. If the latter happens, then the system is back in a fully non-conservative situation. If the former occurs, we stop the simulations and no analysis is performed since it is beyond the scope of this work.

In the case fully non-conservative mass transfer operates, the consequential angular momentum loss caused by material leaving the white dwarf is taken into account as in \citet{King+Kolb-1995ApJ...439..330K} ({\sc use\_other\_extra\_jdot}). This prescription assumes the mass lost through nova eruptions leaves the system with the specific angular momentum of the white dwarf. It is well possible that additional angular momentum loss is generated either through friction between the secondary 
and the expelled envelope \citep{Schreiber-etal-2016MNRAS.455L..16S} or, potentially more likely, through an incomplete ejection which might lead to a common envelope like evolution \citep{Nelemans-etal-2016ApJ...817...69N,Shen-etal-2022arXiv220506283S}. 
This potential angular momentum loss can explain several otherwise inexplicable characteristics of the observed sample of cataclysmic variables \citep{Schreiber-etal-2016MNRAS.455L..16S}. 
However, we are here mostly interested in the early evolution of cataclysmic variables with early type secondaries while
empirical consequential angular momentum loss has been calibrated for cataclysmic variables with low mass secondaries \citep[see][for details]{Schreiber-etal-2016MNRAS.455L..16S}.

Here, we are interested in understanding the evolution of \tar\ which could have have experience thermal timescale mass transfer in the past, and in estimating to what degree previously experienced high mass transfer rates are needed to explain the characteristics of the system we observe today. Such high mass transfer rates are facilitated if the mass ratio reaches values of $q\gtrsim1-1.5$ for both the pseudo-conservative and fully non-conservative cases (see Figure~\ref{fig:Qcrit}). We computed two grids of simulations based on the white dwarf mass determined for \tar\ in Section~\ref{sec:fits}:  

\begin{enumerate}
    \item GRID 1: We considered the white dwarf as a point source of a fixed mass of $0.83\,\Msun$, which is consistent with the mean mass of white dwarfs in cataclysmic variables \citep{Zorotovic-2011A&A...536A..42Z, Pala-2020MNRAS.494.3799P} and our results with free reddening (Table~\ref{tab:parameters}). We explored initial masses for the donor larger than the white dwarf mass, and different initial orbital periods. In general, we explored masses between $M_{\mathrm{donor}}=0.9-2.0\,\Msun$ and initial orbital periods between $P_{\mathrm{orb,i}}=1-5$\,days. At first, we computed a coarsely spaced MESA grid (step of 0.25 \Msun\ and 1 days) and increased the resolution (step of 0.1 \Msun\ and 0.1 days) when the tracks approached the measured parameters of \tar. 
    
    \item GRID 2: We consider a white dwarf mass of \Mwd=0.6\,\Msun\, (Table~\ref{tab:parameters}). The range of donor masses that we considered was $M_{\mathrm{donor}}=0.9-2.0M_{\mathrm{\odot}}$ (in steps of 0.1~\Msun). The initial orbital periods finely sample a range of $P_{\rm orb,i}=2.4-3.5$~days (steps of 0.1~days). For initial orbital periods of $P_{\rm orb,i}<2.4$~days, the accretion rate is still at its maximum value when the orbital period reaches the observed orbital period of \tar\ and hence, they are discarded.
\end{enumerate}

\subsection{Results of the MESA simulations}
\label{sec:}

\begin{figure*}
    \centering
    \includegraphics[width=2.1\columnwidth]{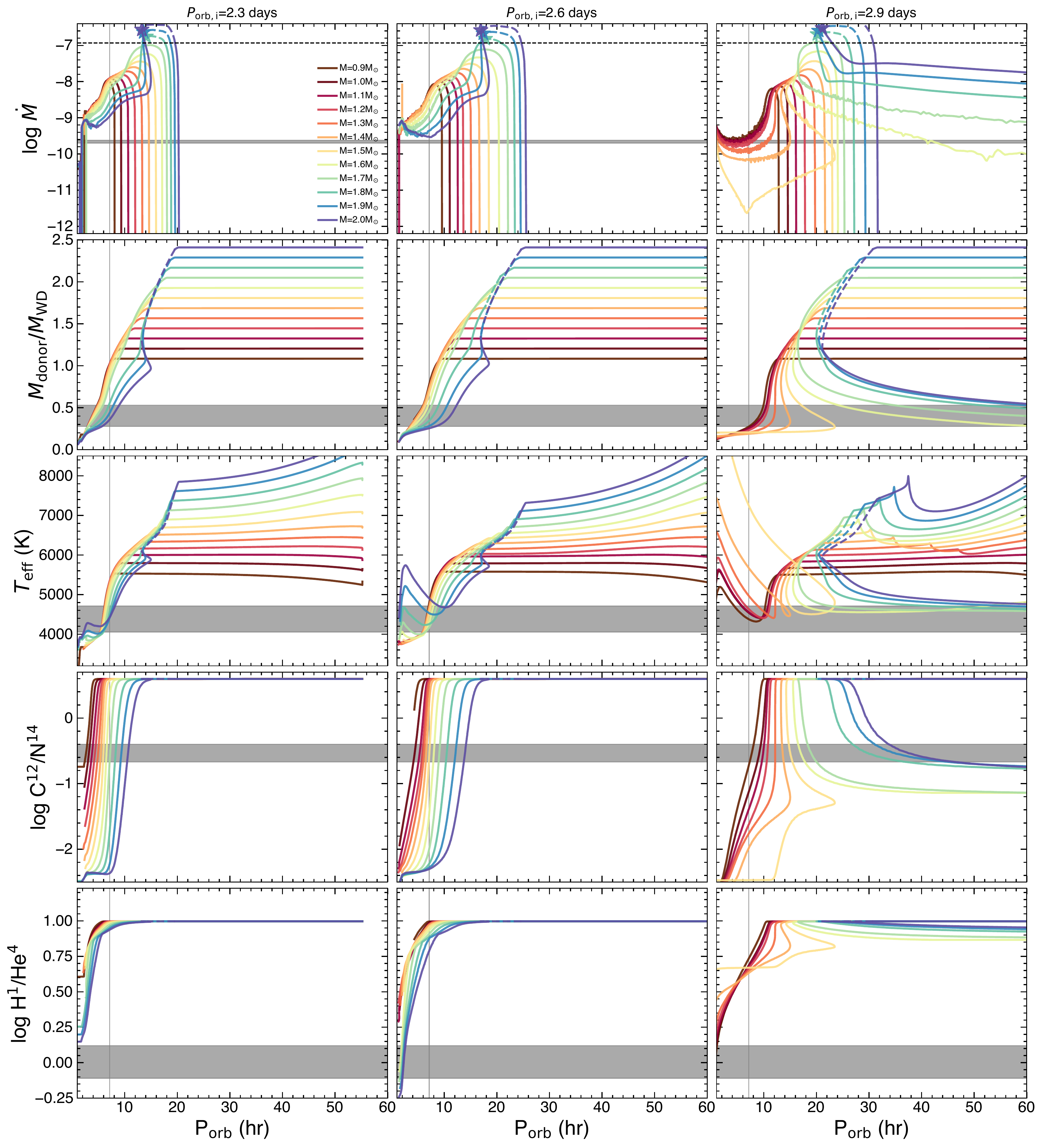}
    \caption{MESA evolutionary tracks assuming pseudo-conservative mass transfer when and if the accretion rate surpasses the hydrogen burning limit (black dashed line; $M_{\rm HB}=1.17\times10^{-7}$~\Msun/yr; \citealt{Wolf-etal-2013ApJ...777..136W}), and fully non-conservative mass transfer otherwise. The star symbols demarcate the upper hydrogen-burning limit. The initial orbital periods of the tracks are shown on the top of the panels, the mass of the white dwarf is fixed to \Mwd=0.83\,\Msun, and the mass of the donor is color coded ($M_{\rm donor}$=0.9-2.0\,\Msun). From top to bottom the panels show the accretion rate, the mass ratio, the donor's effective temperature, the carbon-to-nitrogen ratio, and the hydrogen-to-helium ratio. The gray bands show the measured values for \tar. The observed values of accretion rate, carbon-to-nitrogen ratio, and hydrogen-to-helium ratio are taken from the spectroscopic fits with free reddening (Table~\ref{tab:parameters}). The vertical gray line represents the orbital period for \tar\ \citep{Rodriguez-Gil-etal-2009A&A...496..805R}.  
    \label{fig:MESA_Mwd083}}
\end{figure*}

\begin{figure*}
    \centering
    \includegraphics[width=2.1\columnwidth]{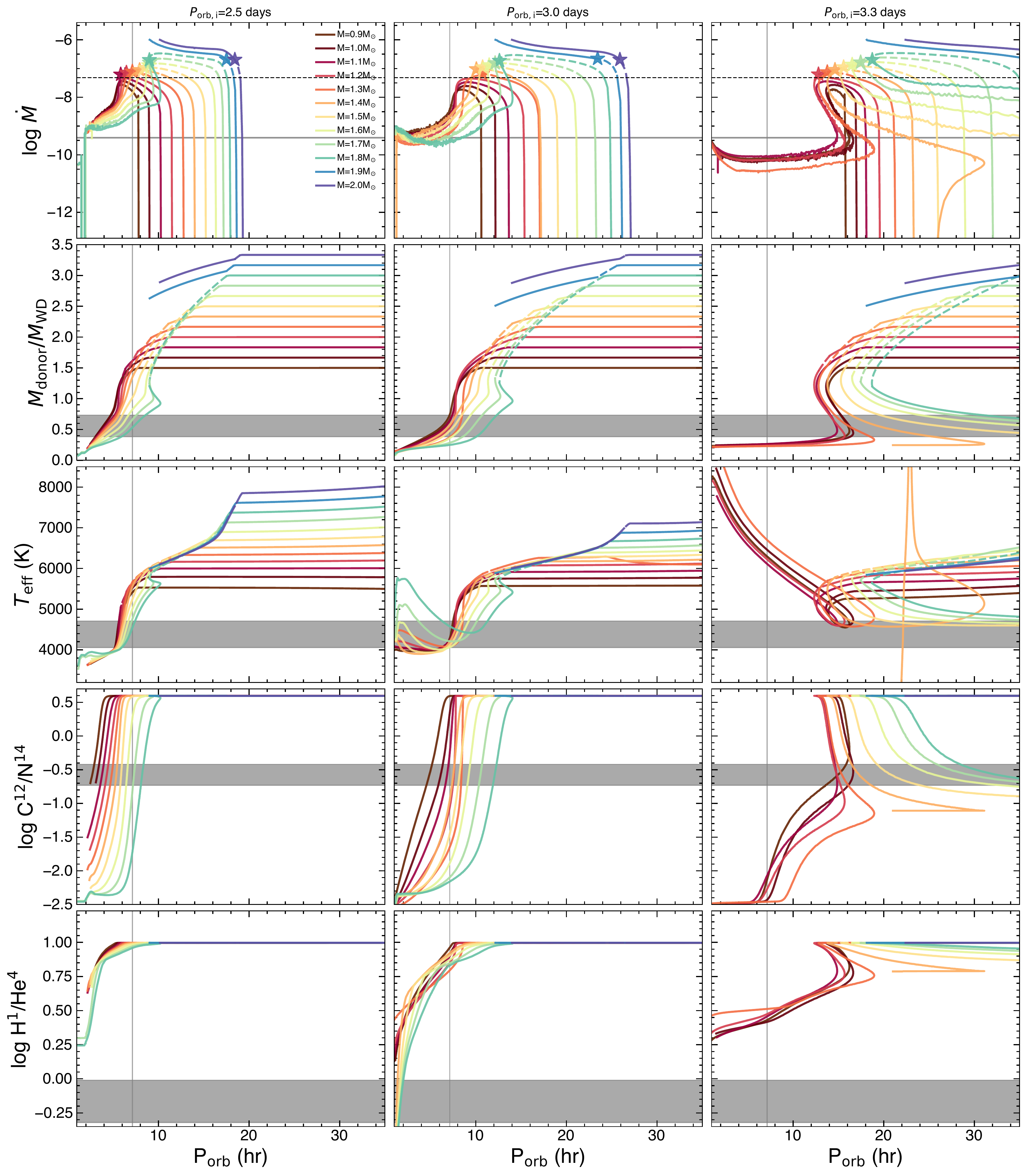}
    
    \caption{Same as in Figure~\ref{fig:MESA_Mwd083} for an initial mass of the white dwarf of $\Mwd=0.6\,\Msun$. The observed values of accretion rate, carbon-to-nitrogen ratio, and hydrogen-to-helium ratio (represented by gray bands) are taken from the spectroscopic fits assuming a fixed reddening from Table~\ref{tab:parameters}. The black dashed line in the top panel corresponds to the lower hydrogen-burning limit for a white dwarf mass of $\Mwd=0.60~\Msun$ ($M_{\rm HB}=0.48\times10^{-7}\,\Msun$/yr, \citealt{Wolf-etal-2013ApJ...777..136W}).
    \label{fig:MESA_Mwd060}}
\end{figure*}

A subset of our MESA simulations is shown in Figure~\ref{fig:MESA_Mwd083} (\ref{fig:MESA_Mwd060}) for GRID~1 (GRID~2). A few tracks evolve off the main sequence, drastically increasing the effective temperatures of the donor star. These systems with initial orbital periods around 2.9 (3.2) days and with the mass of the donor larger than 1.5\,\Msun, evolve towards long orbital periods. The limiting period, known as the bifurcation period, corresponds to the initial period at which a binary configuration (initial masses) will either converge or diverge \citep[e.g.][]{Pylyser-1989A&A...208...52P}. The diverging tracks will eventually detach to form wide systems. 

The first rows of Figure~\ref{fig:MESA_Mwd083} and Figure~\ref{fig:MESA_Mwd060} show the accretion rate as a function of the orbital period. For systems with longer initial orbital periods, the donor stars are at a more advanced evolutionary state at the onset of mass transfer. These evolved donors can start to transfer mass at slightly longer orbital periods. The accretion rates are proportional to the mass of the donor, i.e. the more massive the star, the higher the accretion rate. 

\begin{figure}
    \centering
    \includegraphics[width=0.95\columnwidth]{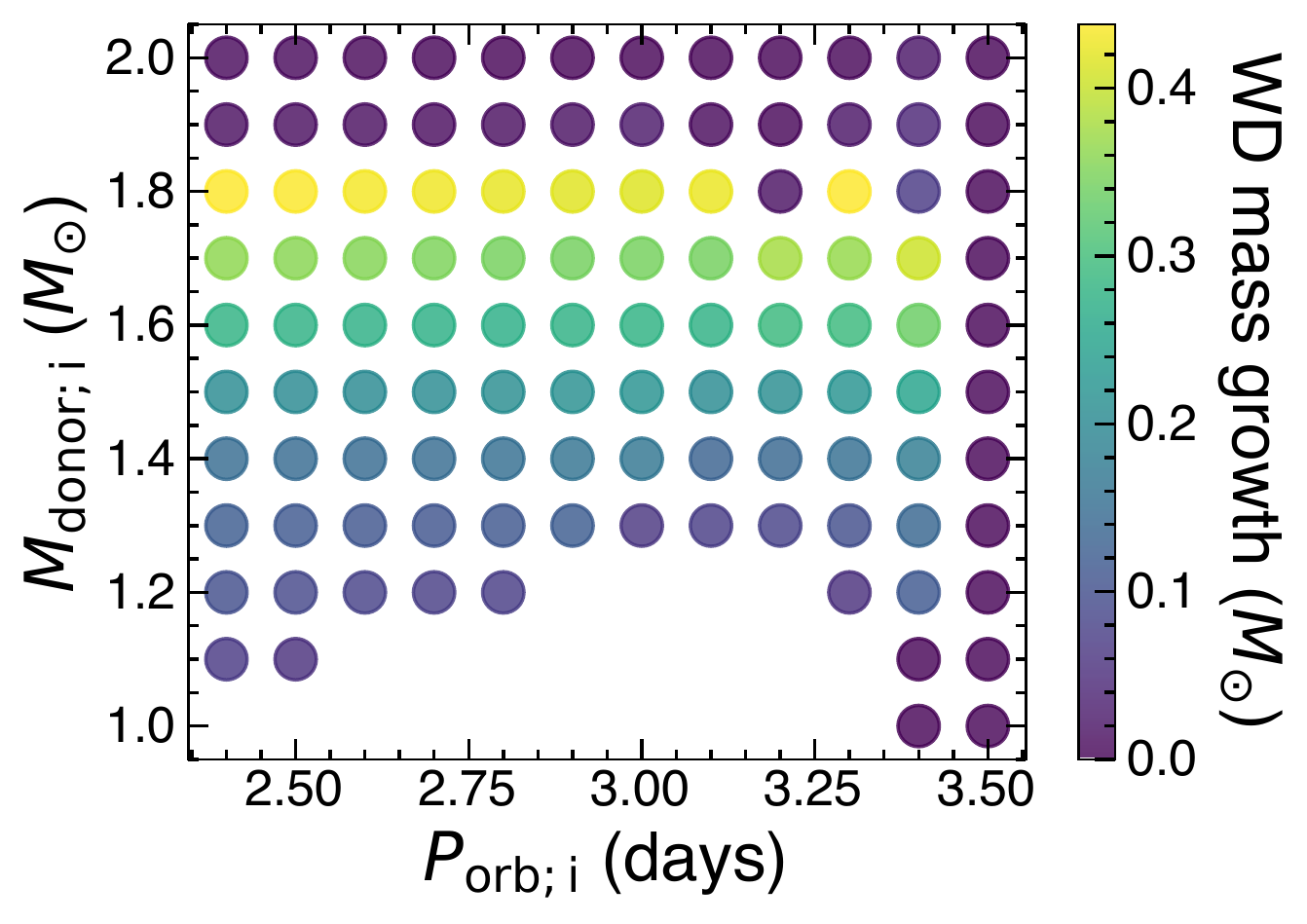}
    \caption{Mass growth ($\Mwd_{;f}-\Mwd_{;i}$) for an initial white dwarf mass of $\Mwd_{;i}=0.6\,\Msun$. A retention of 90\% of the accreted mass is assumed. The configuration that can reach the highest white dwarf mass of $\Mwd\simeq1.04\,\Msun$ corresponds to the grid point $P_{\rm orb,i}=2.4$ and $M_{\rm donor}=1.8$\,\Msun, which is considerable lower than the Chandrasekhar mass limit \citep{Chandrasekhar-1935MNRAS..95..207C}}
    \label{fig:mass_growth}
\end{figure}

Most of the evolutionary tracks of grid\,1 fall below the critical stability boundary of hydrogen burning (for a white dwarf mass of 0.83\,\Msun\ this limit corresponds to $M_{\rm HB}=1.17\times10^{-7}$~\Msun/yr; dashed black lines in the panels in the first row in Figure~\ref{fig:MESA_Mwd083}; \citealt{Wolf-etal-2013ApJ...777..136W}). In contrast, the majority of the tracks calculated for grid~2 reach the hydrogen-burning limit ($\dot{M}_{\mathrm{HB}}=0.48\times 10^{-7}$\,\Msun/yr which is demarcated by a dashed black line in figure~\ref{fig:MESA_Mwd060}). For grid~2, we compute the mass growth of the white dwarf (assuming 90 per cent of the transferred mass to be accreted), which increases with increasing initial mass of the donor (Figure~\ref{fig:mass_growth}). 
This mass growth can be higher than 0.4\,\Msun\ (the highest white dwarf mass that is reached in this grid is $\Mwd=1.04\,\Msun$). For the largest initial donor masses ($>1.8-1.9$\,\Msun), the accretion rates continue increasing, even surpassing the upper $3\times\,\dot{M}_{\mathrm{HB}}$ bound (stars in the top panel in Figure~\ref{fig:MESA_Mwd060}), evolving towards dynamically unstable mass transfer and potentially common envelope evolution, significantly limiting the mass growth of the white dwarf during the stable hydrogen burning phase. For lower initial donor masses ($<1.8-1.9$\,\Msun), the accretion rate reaches a maximum and then drastically decreases when the mass ratio approaches unity, causing a slight increase in the orbital period.

The mass ratios, shown in the second row in Fig.~\ref{fig:MESA_Mwd083} and Fig.~\ref{fig:MESA_Mwd060}, decrease towards shorter orbital periods due to mass loss of the donor. In those tracks with larger initial masses of the donor, the systems approach the critical mass ratio ($q_{\mathrm{crit}}$) for dynamical stability of mass transfer. 
Figure~\ref{fig:Qcrit} shows a few of our MESA tracks from grid~1 (stars) and grid~2 (squares), with initial masses of the donors corresponding to 1.0\,\Msun\ (blue), 1.4\,\Msun\ (red), and 1.9\,\Msun\ (brown). All systems undergoing dynamically stable mass transfer evolve from top-right to bottom left in the $q-M_{\mathrm{donor}}$ parameter space. For fully non-conservative mass transfer, the slope of the evolutionary track is less steep than for pseudo-conservative mass transfer. Some systems can experience non-conservative thermally unstable mass transfer (e.g. \Mwd=0.83\,\Msun\ plus $M_{\mathrm{donor}}=1.4-1.7\,\Msun$, $P_{\rm orb,i}=2.y$~days).


For the largest mass ($M_{\rm donor}=1.9$~\Msun; brown square in Figure~\ref{fig:Qcrit}), the system can experience a conservative mass transfer is which is dynamically unstable given the high accretion rate (top middle panel in Figure~\ref{fig:MESA_Mwd060}).

For $M_{\rm donor}=1.4$~\Msun\ the systems are initially unstable against thermal time scale mass transfer and reach the hydrogen burning limit (top middle panel in Figure~\ref{fig:MESA_Mwd060}). During the short pseudo conservative mass transfer phase, the mass ratio decreases quickly, and the systems reach thermally stable mass transfer, which coincides with dropping below the hydrogen burning limit. Hence, the evolution continues as the non-conservative thermally stable mass transfer becoming a CV with an evolved low-mass donor.

Finally, the track with $M_{\rm donor}=1.0$~\Msun\ (blue) is thermally stable and the accretion rate remains below of the hydrogen burning limit.


The effective temperature of the donor is shown in the third row in the panels in Figure~\ref{fig:MESA_Mwd083} and Figure~\ref{fig:MESA_Mwd060}. Given that the mass of the donor is decreasing due to mass transfer, the effective temperature also decreases. However, Figure~\ref{fig:MESA_Mwd083} shows that some tracks reach a minimum and then a subsequent increase of the effective temperature with decreasing period. This behavior becomes more pronounced for the largest initial donor masses. For example, the effective temperature for $M_{\mathrm{donor}}=2.0$\,\Msun\ (blue track in Figure~\ref{fig:MESA_Mwd083}) that has initial orbital period of $P_{\mathrm{orb,i}}=2.6$\,days reaches this minimum (\Teff$\sim$4700\,K) when the orbital period has decreased to around 10\,h. Then the effective temperature increases, until at around $P_{\mathrm{orb}}\simeq2.4$\,hr, the effective temperature reaches its peak at \Teff$\simeq$5700\,K, then subsequently starts to decrease again. This behavior could be explained by the surface convective layer reaching the energy generation layer, which for more massive initial masses, has moved outward. The chemical profile of the core is mainly helium, with a thin hydrogen layer. Thus, if these donors are able to reach high temperatures they could eventually turn into ELM white dwarfs \citep{El-Badry-etal-2021MNRAS.505.2051E, El-Badry-etal-2021MNRAS.508.4106E}.

The fourth and fifth rows in the panels in Figure~\ref{fig:MESA_Mwd083} and Figure~\ref{fig:MESA_Mwd060} show the surface carbon-to-nitrogen (\cno) ratios and the hydrogen-to-helium ratios (\HeH), respectively. They present a steep decrease once the mass of the donors have been eroded down to less than $\sim 0.8\,\Msun$. A consequence of the convective regions reaching deeper layers in the donor star is the dredge up of material processed through the CNO cycle. As a consequence, the carbon-to-nitrogen ratio at the donor's surface inverts \citep{Schenker-etal-2002MNRAS.337.1105S}. Similarly, the surface is enriched in helium due to convection reaching inner regions.


\begin{figure*}
    \centering
    \includegraphics[width=1\columnwidth]{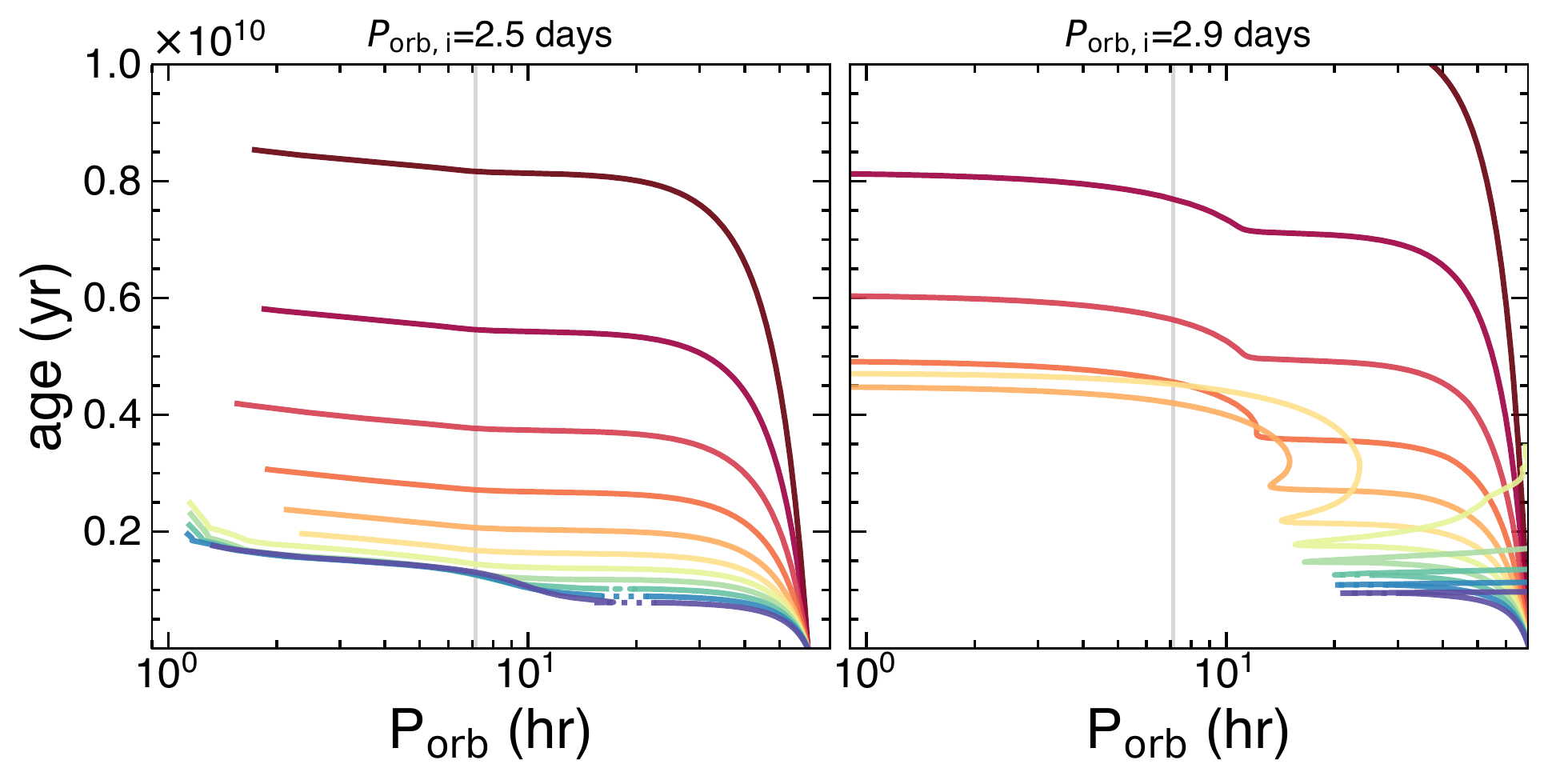}
    \includegraphics[width=1\columnwidth]{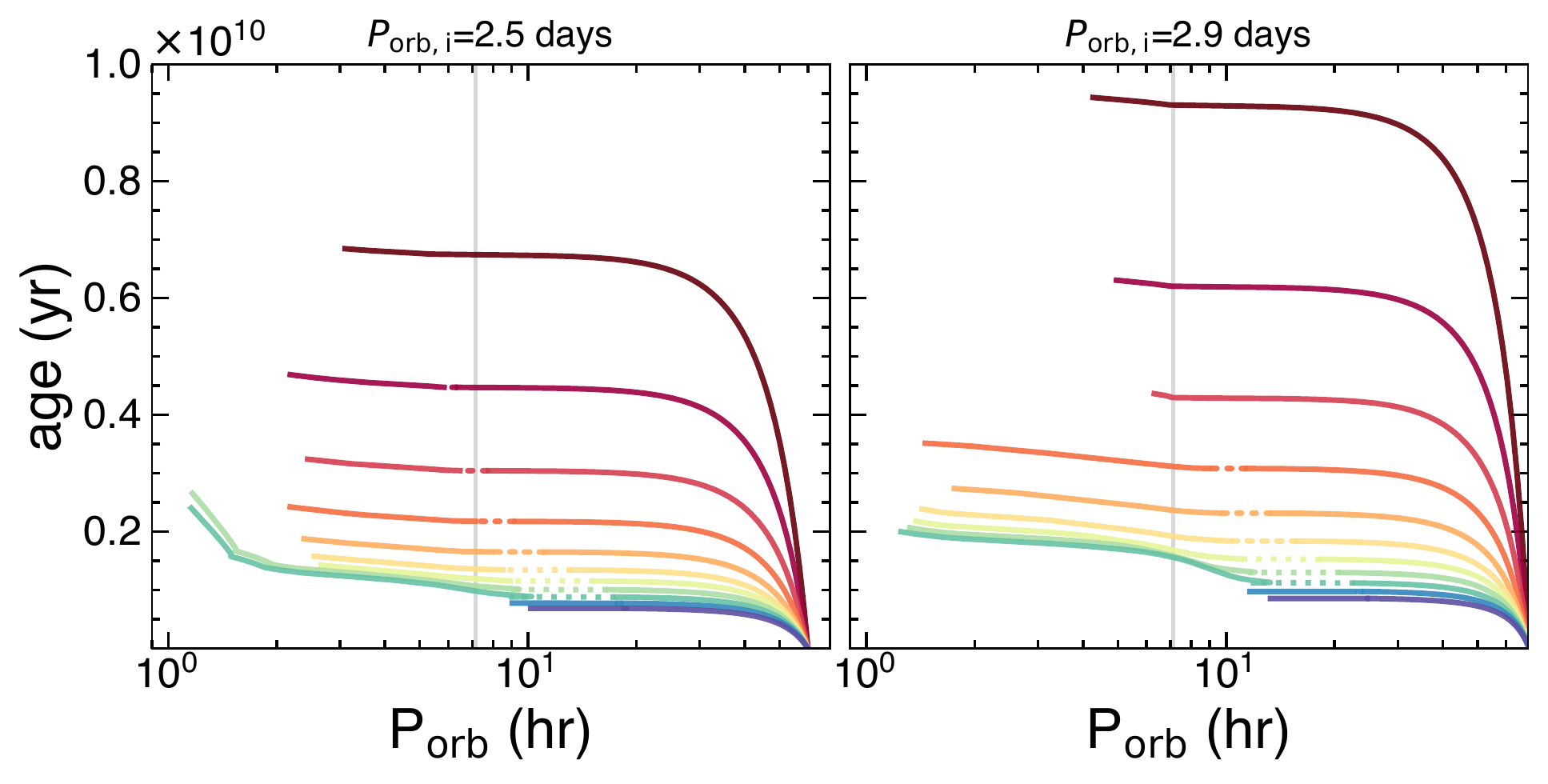}
    \caption{Age versus orbital period for grid of $\Mwd=0.83\,\Msun$ (left) and $\Mwd=0.83\,\Msun$ (right). The colors represent the initial mass of the donor in the simulations coded as in Figure~\ref{fig:MESA_Mwd083}. The initial orbital period is shown on the top of each panel. The gray vertical lines shows the orbital period of \tar. 
\label{fig:age-porb}}
\end{figure*}

For completeness, also shown is the age of the donor as a function of the orbital period is plotted in Figure~\ref{fig:age-porb}, which shows that the evolution is faster for those configurations having larger donor star masses. For the largest masses (blueish tracks), the orbital period shrinks to less than 15 hours in only $1-2$\,Myr, and this time increases towards larger initial orbital periods. With mass transfer, the donor mass goes down as it develops a convective envelope that deepens with the decreasing mass. As a consequence, the evolution slightly slows down. This effect is more noticeable when the initial orbital periods are longer.

\section{Past evolution of HS0218+3229}
\label{sec:past_evolution}

The binary parameters combined with the stellar parameters of the donor can be used to distinguish which evolutionary MESA simulation can describe the measured parameters of \tar. These parameters (Table~\ref{tab:parameters}) correspond to (1) orbital period of \tar\ \citep{Rodriguez-Gil-etal-2009A&A...496..805R}, (2) accretion rate (this work; section \ref{sec:results_fits}), (3) carbon-to-nitrogen ratio (this work, section \ref{sec:results_fits}), (4) the mass of the donor \citep{Rodriguez-Gil-etal-2009A&A...496..805R}, and (5) the spectral type of the donor \citep{Rodriguez-Gil-etal-2009A&A...496..805R} as proxy of its effective temperature. 
Based on comparison templates and infrared colours, \citet{Rodriguez-Gil-etal-2009A&A...496..805R} ruled out an M-type for the donor, favoring a K4–K5 dwarf donor. They conclude that the spectral type is likely to be K5 based on the strength of the TiO bands in their optical spectra. Indeed, the broad feature around $\sim5200$\,\AA\ (which is the superposition of the MgH at 5180\,\AA, a TiO band at $\lambda\lambda4954-5200$, and the \Ion{Mg}{i} triplet at$\lambda\lambda5168-5185$) is a typical feature in K-type stars \citep{Rodriguez-Gil-etal-2009A&A...496..805R}. We convert spectral type into effective temperature by using the sample of eclipsing binaries for which their stellar parameters have been determined \citep{Torres-2010A&ARv..18...67T, Southworth-2015ASPC..496..164S}\footnote{https://www.astro.keele.ac.uk/jkt/debcat/}. We extracted a sub-sample of all stars with spectral types within K4-K6 (to account for some uncertainty), resulting in a small sample of seven stars. The mean and standard deviation of the effective temperature for this sample corresponds to $4382\pm326$\,K, which is the value adopted for the donor in \tar.

We use an interpolation method based on Gaussian processes (GP) to determine the region of the MESA grids that best describes the observed parameters of \tar. Using the parameters marked with a star in Table \ref{tab:parameters}, we adopted two methods where (1) we interpolate the values of each parameter independently to find the region of the MESA grid for which the values are the closest to the observed parameters, and (2) we interpolate and minimise an averaged-weighted error to find the point in the MESA grid closest to the observed parameters. Although the second method finds an optimal solution, we investigate the sensitivity of each parameter using the first method.

\subsection{Method 1: Combining Gaussian processes on the individual parameters}
\label{subsec:GP}

\begin{figure*}
    \centering
    \includegraphics[width=1\columnwidth]{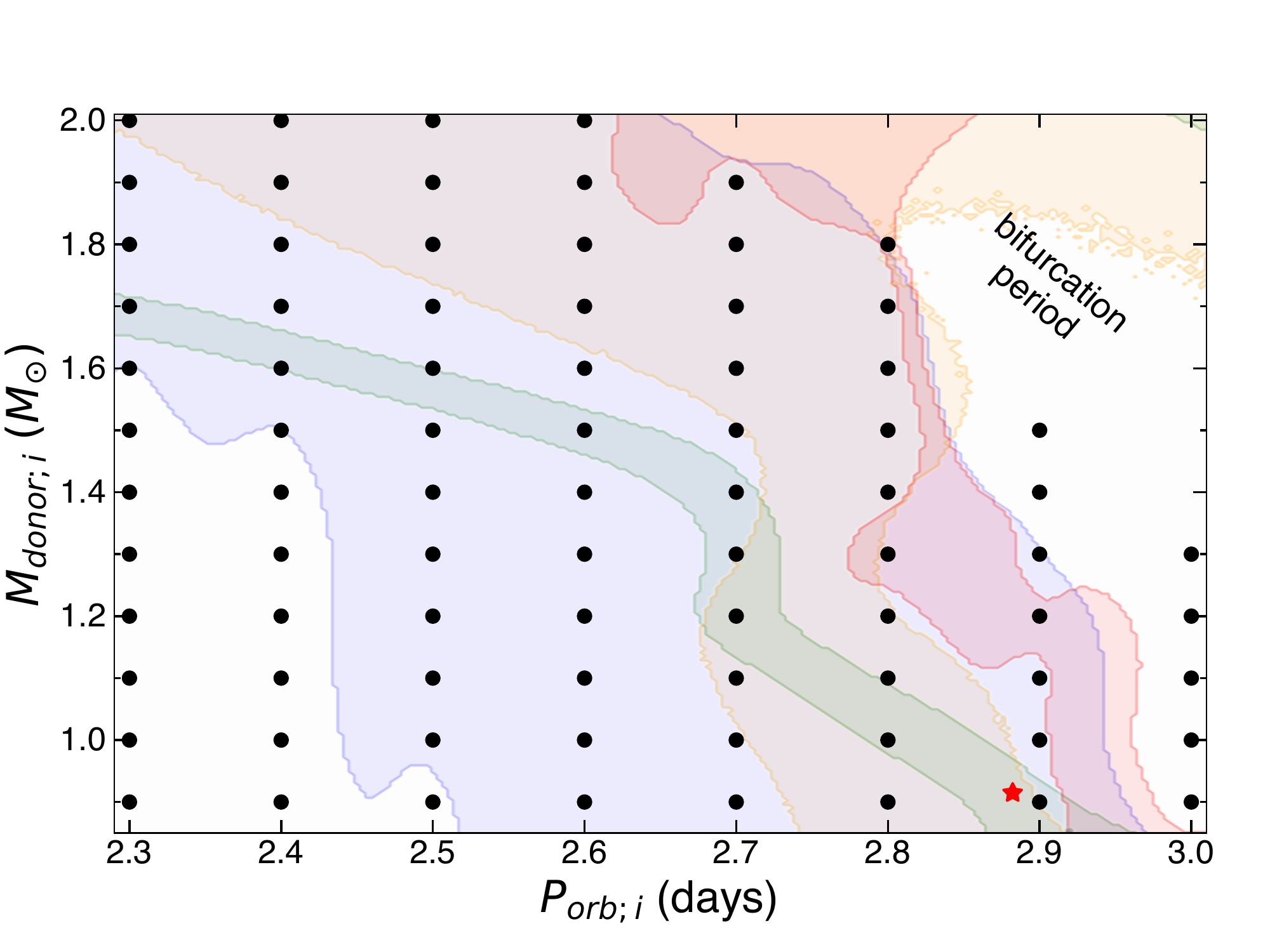}
    \includegraphics[width=1\columnwidth]{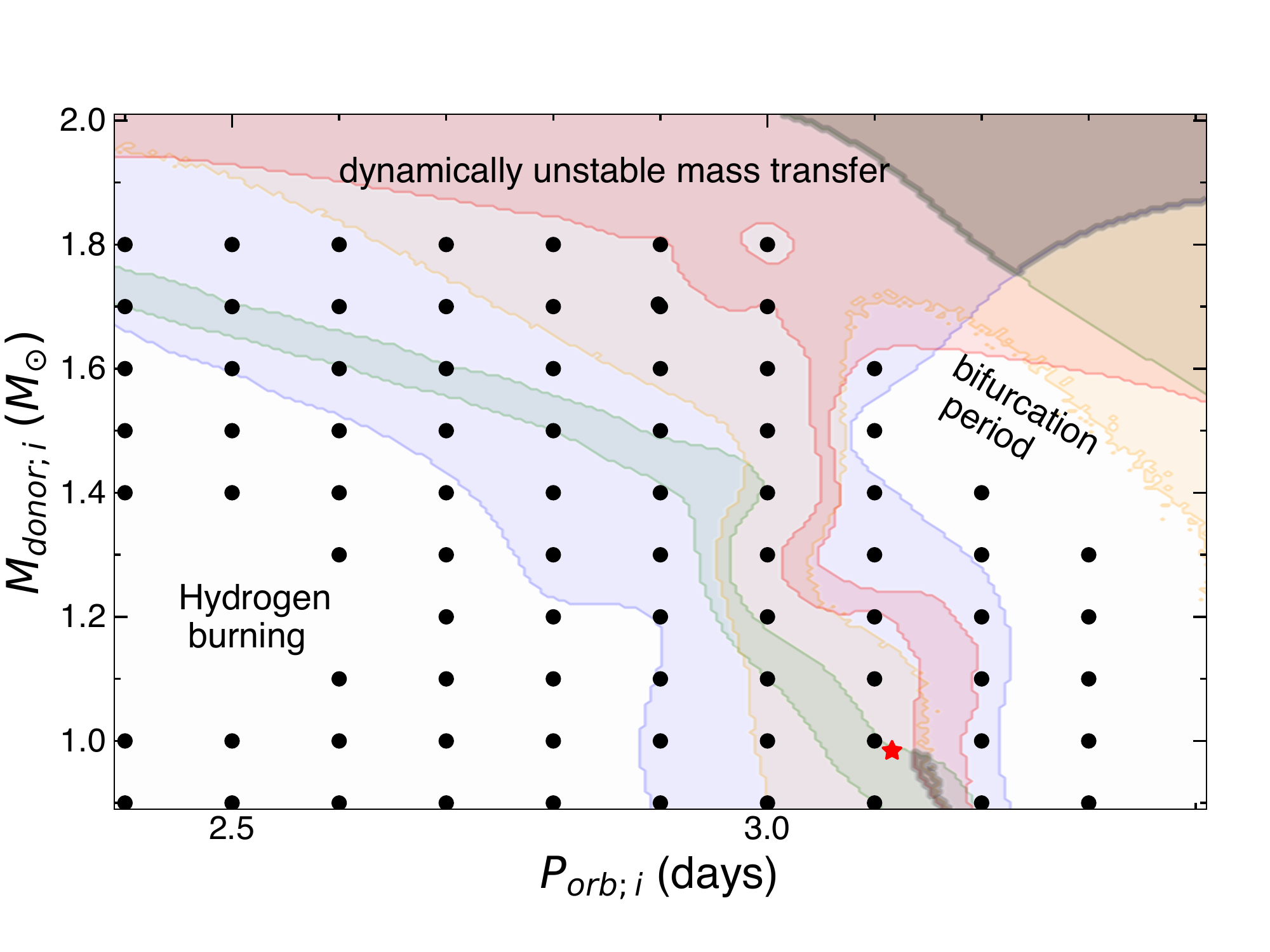}
    \caption{Best solutions of the MESA grid (black dots) that reproduces the observed parameters of \tar\ for when $\Mwd=0.83\,\Msun$ (left) and  $\Mwd=0.60\,\Msun$ (right) cases. Two methods were used, the first one is Gaussian process fits to the effective temperature (blue), the carbon-to-nitrogen ratio (green), the donor's mass (orange), and the accretion rate (red), where the overlapping region (gray region close to the red star) represents the best solution (which only occurs in right panel, and solution below 0.93\,\Msun\ requires times slightly longer than the age if the universe). The second method is Gaussian process fit to an averaged-weighted error (see section~\ref{subsec:weighted-error}) to identify the minimum (red star) in the parameter space. 
    \label{fig:gp+error}}
\end{figure*}

For each of the parameters (marked with a star in Table~\ref{tab:parameters}), a GP is fitted to interpolate the values of the parameter at \tar's orbital period (Table~\ref{tab:parameters}) for all points in the MESA grid. The interpolation estimates the value of the parameter for orbital periods between $P_{\mathrm{orb,i}}=2.4-3.4$\,days and masses between $M_{\mathrm{donor}}=0.9-2.0M_{\mathrm{\odot}}$ for when $\Mwd=0.60\,\Msun$, and orbital periods between $P_{\mathrm{orb,i}}=2.3-3.0$\,days and masses between $M_{\mathrm{donor}}=0.9-2.0M_{\mathrm{\odot}}$ for the case when $\Mwd=0.83\,\Msun$. Then, the interpolated surface is compared to the observed parameter to determine the region of the MESA grid that best describes the observed parameter. Points in the optimal region are denoted as \textit{fit points}. The GP is implemented using a Matérn covariance function with smoothness parameter $\nu=3/2$ and is implemented using the \textsc{gpflow} package from \textsc{python} \citep{GPflow2017}. The method is described in detail in Appendix \ref{sec:appendix_method1}. Results are shown in Table~\ref{tab:result_sims} and Figure~\ref{fig:gp+error}, which show the regions of fit points for the effective temperature (blue), the carbon-to-nitrogen ratio (green), the donor's mass (orange), and the accretion rate (red). While the effective temperature provides the weakest constraint, the carbon-to-nitrogen ratio is the parameter that provides the strongest constraint. Only for the case of $\Mwd=0.60\,\Msun$, there exist an overlapping region (grey), which correspond to the best initial configurations of \tar\ for this method, which manifests a very unsmoothed surface of the solution.

There are regions with no gridpoints (black dots) in Figure~\ref{fig:gp+error}. In the right panel, the top region (initial donor masses $M_{\rm donor;i}\geq1.9$~\Msun) demarcates configurations where the evolution undergoes dynamically unstable mass transfer. A second region near the bottom left, these missing points represent evolution tracks where hydrogen burning is taking place at the orbital period of \tar. Finally, a diagonal region towards long initial orbital periods exists, which corresponds to the bifurcation periods, and thus these tracks never reach the observed orbital period of \tar. In contrast the left panel shows absence of gridpoints due the the bifurcation period.

In order to discern how the measurable parameters evolve for binary configurations enclosed by this overlapping region we randomly selected five data points, that sample the solution region (gray region in right panel of Figure~\ref{fig:gp+error}). These tracks are shown in Figure~\ref{fig:future} as orange lines. The evolution for these additional tracks was stopped when the donor's mass reached 0.06\,\Msun\ (i.e. the mass limit was stretched down to 0.06\,\Msun\ just below the hydrogen burning limit; \citealt{Dieterich-etal-2014AJ....147...94D}), where the stellar properties can be considered valid for the same MESA inlist.

The accretion rate is shown in the top panel of Figure~\ref{fig:future}. Whereas the predicted accretion rate for $\Mwd=0.60\,\Msun$ (orange solid lines) agrees with our estimate of the accretion rate (orange dot; fixed reddening in Table~\ref{tab:parameters}). 
The effective temperature (second panel) agrees with the prediction. The donor mass (third panel) and carbon-to-nitrogen ratio (fourth panel) are predicted to be slightly lower than the measured for the tracks. For completeness, we also show the hydrogen-to-helium ratio which clearly disagrees with the prediction, however this quantity was not considered during the GP fits since it is not reliable. 

None of the tracks reach the accretion rate for stable hydrogen burning. Thus, as the white dwarf mass remains constant along the evolution, the mass values assumed in the MESA simulations are consistent with the value measured for the mass of the white dwarf in \tar\ (Table~\ref{tab:parameters}). However, these five tracks for $\Mwd=0.60\,\Msun$ tracks undergo a phase of thermal timescale mass transfer.

\subsection{Method 2: Gaussian process to the averaged weighted error}
\label{subsec:weighted-error}

A GP is fitted to interpolate and minimise an averaged-weighted error to find the point in the MESA grid that best describes the observed parameters. The error is defined based on the parameters of the MESA evolutionary tracks at the observed orbital period (Table~\ref{tab:parameters}). Particularly, for a point $(P_{\mathrm{orb}}, M_{\mathrm{donor}})$ in the MESA grid, the error is defined as:

\begin{equation}
\centering
\epsilon(P_{\mathrm{orb}}, M_{\mathrm{donor}}) = \frac{1}{4}\sum_{i=1}^{4} \int g_i(p)\frac{|x_i(P_{\mathrm{orb}}, M_{\mathrm{donor}}) - p|}{\sigma_i}dp,   
\label{eq:error}
\end{equation}

\noindent where $i$ is the index of the parameter (accretion rate, the mass of the donor, the donor's effective temperature, the carbon-to-nitrogen ratio), $g_i(p)$ is the empirical distribution of the observed values of the parameter $i$, $x_i$ is the value of $i$ at $(P_{\mathrm{orb}}, M_{\mathrm{donor}})$ and $\sigma_i$ is the standard deviation of the value of $i$ for all the MESA grid points. The \textit{best track} is defined as the initial orbital period $P_{\mathrm{orb}}$ and the initial mass of the donor $M_{\mathrm{donor}}$ that minimises the mean of the marginal distribution of the GP at the point $(P_{\mathrm{orb}}, M_{\mathrm{donor}})$. 

The GP is implemented using the same specifications as for Method 1 in Section \ref{subsec:GP}. The method is described in detail in Appendix \ref{sec:appendix_method2}. Results for the best track are P$_{\mathrm{orb;i}}=2.88$\,days and $M_{\rm donor;i}=0.91\,\Msun$ for $\Mwd=0.83\,\Msun$ and P$_{\mathrm{orb;i}}=3.12$\,days and $M_{\rm donor;i}=0.98 \Msun$ for $\Mwd=0.60\,\Msun$ case (red stars in Figure~\ref{fig:gp+error}). While these solutions are not fully enclosed by the region found by the previous method (section~\ref{subsec:GP}), the solutions fall very close in the parameter space (Figure \ref{fig:gp+error}). Based on the values of the minimum error (Table~\ref{tab:result_sims}), the results for $\Mwd=0.83\,\Msun$ provides a slightly better solution than $\Mwd=0.60\,\Msun$. 

With the exception of the accretion rate for the $\Mwd=0.83\,\Msun$, the evolution predicted (blue dotted line in Figure~\ref{fig:future}) is similar to the solution for $\Mwd=0.60\,\Msun$ obtained with method 1. In contrast, the best solution for $\Mwd=0.63\,\Msun$ (orange dotted line) slightly diverge from the one found by method 1 (orange solid line). However, the behavior is similar as for method 1, i.e. the effective temperature agrees better with the prediction, the donor mass and the carbon-to-nitrogen ratio are predicted to be lower than the observed ones, and the accretion rate is predicted to be higher than the observed one. Consistent with method 1, none of the tracks undergo hydrogen burning, and  only the track that assume $\Mwd=0.60\,\Msun$ experiences thermal timescale mass transfer.

The two methods lead to consistent solution regions on the parameter space of the MESA grid. Both methods provide different insights, while method 1 shows the sensitivity that each parameter has on the solution, method 2 is able to quantify that the solution for $\Mwd=0.83\,\Msun$ is slightly better than for $\Mwd=0.60\,\Msun$ (see the error in Table~\ref{tab:result_sims}).

\begin{table}
\begin{center}
\caption{Results of best-fit initial configurations for \tar, found assuming a white dwarf mass of $\Mwd=0.83\,\Msun$ (first block) and $\Mwd=0.60\,\Msun$ (second block). These best solution are found by fitting the measured parameters of \tar\ (those demarcated with a star in Table~\ref{tab:parameters}) using two methods (for details see sections~\ref{subsec:GP} and \ref{subsec:weighted-error}). The third column shows the value of $\epsilon$ (eq.~\ref{eq:error}) calculated for the minimum, i.e. the value from column 2. \label{tab:result_sims}}
\begin{tabular}{l c c c}
     \hline\hline
       Parameter  &  Method 1 & Method 2 & $\epsilon$ (eq.~\ref{eq:error}) \\
       \hline
       $M_{\rm donor;i}\,(\Msun)$ & -- & 0.91 & \\
       P$_{\mathrm{orb;i}}$\, (days) & -- & 2.88 & 0.36 $\pm$ 0.02\\
       $\Mwd\,(\Msun)$ & 0.83 & 0.83\\
       \hline
       $M_{\rm donor;i}\,(\Msun)$ & 0.90-0.98  & 0.98 & \\
       $P_{\mathrm{orb;i}}$\,(days) & 3.14-3.16 & 3.12 & 0.45 $\pm$ 0.03\\
       $\Mwd~(\Msun)$ & 0.60 & 0.60 & \\
       \hline
\end{tabular}
\end{center}
\end{table}

\section{Discussion}
\label{sec:discussion}

Based on the solutions that explain the initial configuration post common envelope here we discuss the future of \tar\ as well as the influence of the tertiary around \tar.

\subsection{Future of \tar}
\label{sec:future}

\begin{figure}
    \centering
    \includegraphics[width=0.95\columnwidth]{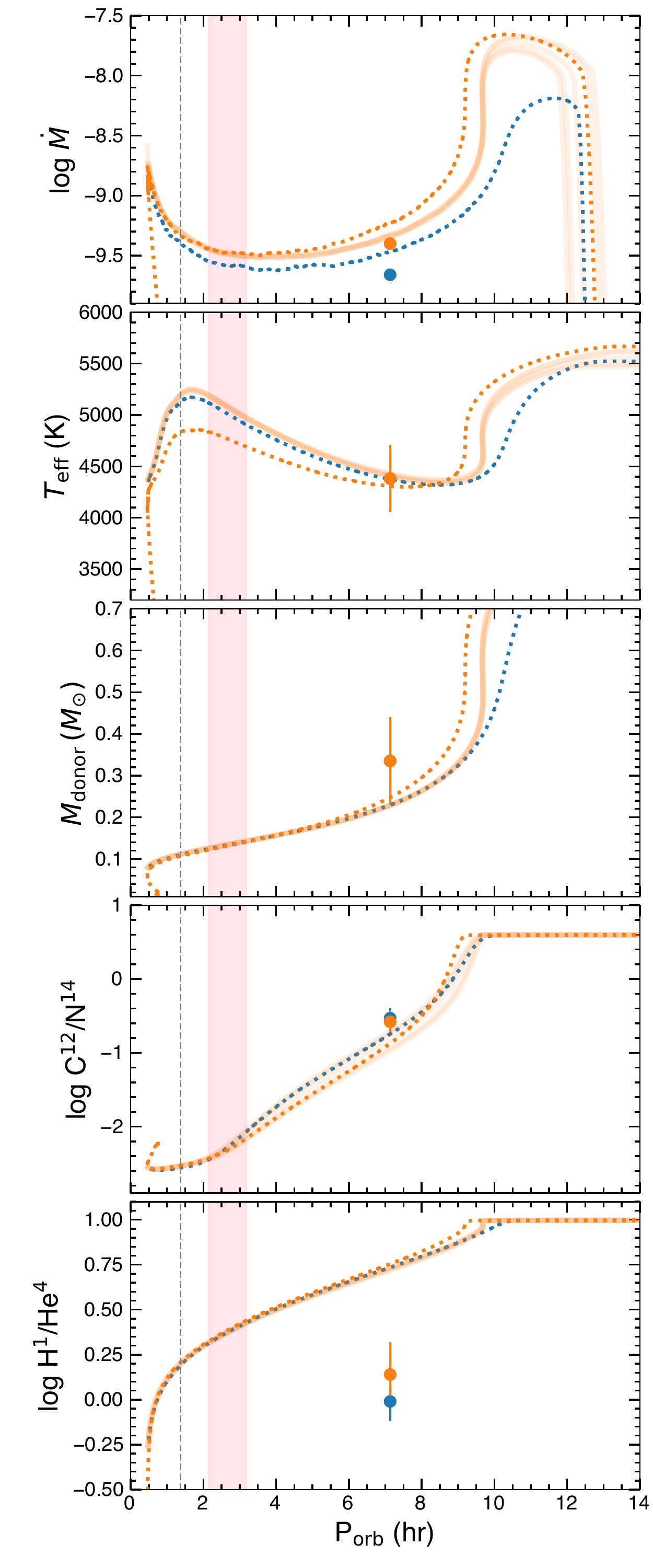}
    \caption{Best evolutionary solutions for $\Mwd=0.83\,\Msun$ (blue) and $\Mwd=0.60\,\Msun$ (orange) found using two methods: minimization of average weighted error (dashed, see section~\ref{subsec:weighted-error}) and combination of Gaussian process fits to the individual parameters (solid lines; see section~\ref{subsec:GP}). From top to bottom are shown the accretion rate, the effective temperature, the mass of the donor, the carbon-to-nitrogen ratio, and the hydrogen-to-helium ratio (we stress that the hydrogen-to-helium ratio is not included in the fits, it is just shown for comparison purposes). The measured values of these parameters for \tar\ are shown with dots obtained from fits assuming free (blue) and fixed (orange) reddening (Table~\ref{tab:parameters}). The tracks continue accreting when crossing the period gap ($2.15-3.18$\,h; pink band), and evolve towards shorter orbital periods than the well-established period minimum of CVs ($71.1\pm1$\,min, vertical gray dashed line).
    \label{fig:future}}
\end{figure}

Based on the solutions that best describe the observed parameters found in Section~\ref{sec:past_evolution}, we explore the future evolution of \tar, which is shown in Figure~\ref{fig:future}. Except for the effective temperature, the future prediction of the parameters is similar for both $\Mwd=0.83\,\Msun$ and $\Mwd=0.60\,\Msun$ cases. For all cases, the effective temperature in the future is predicted to be hotter at shorter orbital periods (with the caveat that the solution found by method 1 and method 2 diverge)

Using the solution from method 2, for the case where the white dwarf mass is assumed $\Mwd=0.60\,\Msun$ ($\Mwd=0.83\,\Msun$) we found that \tar\ started to transfer mass at an orbital period of 1.32 (1.88) days when the system had an age of 10.87 (13.4) Gyr, and the system reaches the current orbital period of \tar, at 11.78 (16.17; note this is longer than the age of the universe) Gyr.

In general, two clear aspects of the evolution can be identified. The first one is that systems continue accreting when crossing the period gap (pink band in Figure~\ref{fig:future}) as found in previous binary population synthesis works \citep[e.g.][]{Podsiadlowski-2003MNRAS.340.1214P}. Second the tracks reach orbital periods ($\sim 30$\,min) that are shorter than the minimum period reached by CVs (i.e. $71.1\pm1$\,min; \citealt{Knigge-2006MNRAS.373..484K}; dashed gray line in Figure~\ref{fig:future}). 

The predicted orbital periods and donor masses that \tar\ can reach are consistent with the parameter space covered by many AM\,CVn stars. AM\,CVn stars are white dwarfs accreting from hydrogen-deficient companions. Three formation channels exist. The first scenario constitutes double white dwarfs that result from two common envelopes, and evolve to ultrashort period by emission of gravitational waves. In the second scenario a white dwarf accretes from a helium-burning star, which could be the remnant core of an evolved star. Similarly to CVs when the donor star becomes semi-degenerate, the orbital period reaches a minimum, and subsequently the orbit starts to expand. The third scenario assumes that AM\,CVn systems can be formed from CVs with evolved donor stars. Hence, \tar\ could be the progenitor of an AM\,CVn system through the evolved donor star channel \citep{Liu-etal-2021ApJ...910...22L}. 

However, the predicted amount of hydrogen left in the donor's envelope is significant when reaching the minimum period of CVs (at 71\,min): 0.67 and 0.15 (in mass fraction) for $\Mwd=0.83\,\Msun$ and $\Mwd=0.60\,\Msun$, respectively. And thus at this stage it can explain systems like V485-Cen \citep[$P_{\rm orb}=59$\,min;][]{Augusteijn-etal-1996A&A...311..889A}, EI\,Psc \citep[$P_{\rm orb}=64$\,min;][]{Thorstensen-etal-2002ApJ...567L..49T}, V418\,Ser \citep[$P_{\rm orb}=64.2$\,min;][]{Kennedy-etal-2015ApJ...815..131K}, and CRTSJ111126.9+571239 \citep[$P_{\rm orb}=56$\,min;][]{Kennedy-etal-2015ApJ...815..131K}. These few short period systems contain evolved donors and show hydrogen in their spectra. Thus, we speculate that \tar\ will join this class of objects. Recently, it has been proposed that the donors in WD+eMS systems could turn into ELM white dwarfs when their temperatures reach values higher than $6500-7000$\,K and the system detach due to supressing magnetic braking \citep{El-Badry-etal-2021MNRAS.505.2051E, El-Badry-etal-2021MNRAS.508.4106E}, however our simulation tracks show that the systems continue with mass transfer since magnetic braking is operating even when the effective temperature of the donor significantly increases. However, our MESA simulations show that the donor in \tar\ will never reach such temperature (Figure~\ref{fig:future}). Therefore, the system once experiemce the period bounce can become an AM\,CVn star, as Gaia14aae which has been proposed to have formed through this channel \citep{Green-etal-2018MNRAS.476.1663G}.

\subsection{Third object as a perturber?}

Gaia data reveal the existence of an object (Gaia DR2 325051817976249600) with a similar parallax ($\varpi=1.92 \pm 0.25$) and proper motion ($\mu_{\rm RA} = -10.33 \pm 0.64$ and $\mu_{\rm dec} =	-13.30	\pm 0.58$) to \tar\ ($\mu_{\rm RA} =-10.85\pm0.10$ and $\mu_{\rm dec} = -13.40 \pm	0.10$). This object is located at a projected distance of $3000-3500$\,AU from \tar, is bright in optical and infrared images (e.g PanSTARRS $g$' and $z$', and \textit{2MASS} bands), but undetected in ultraviolet images (\textit{GALEX}). With a \textit{Gaia} absolute magnitude of $\simeq$ 10.31\,mag, the object sits on the M-type main sequence stars in the \textit{Gaia} color-magnitude diagram, thus it is likely that this object has a stellar nature. The chance alignment probability is very small, i.e. $\mathcal{R}= 0.000157$ \citep{El-Badry-etal-2021MNRAS.506.2269E}, which supports that the star is gravitationally bound. This triple forms a hierarchical system, hence we here investigate if this third body can perturb the inner binary during the CV evolution. 

\begin{figure}
    \centering
    \includegraphics[width=1\columnwidth]{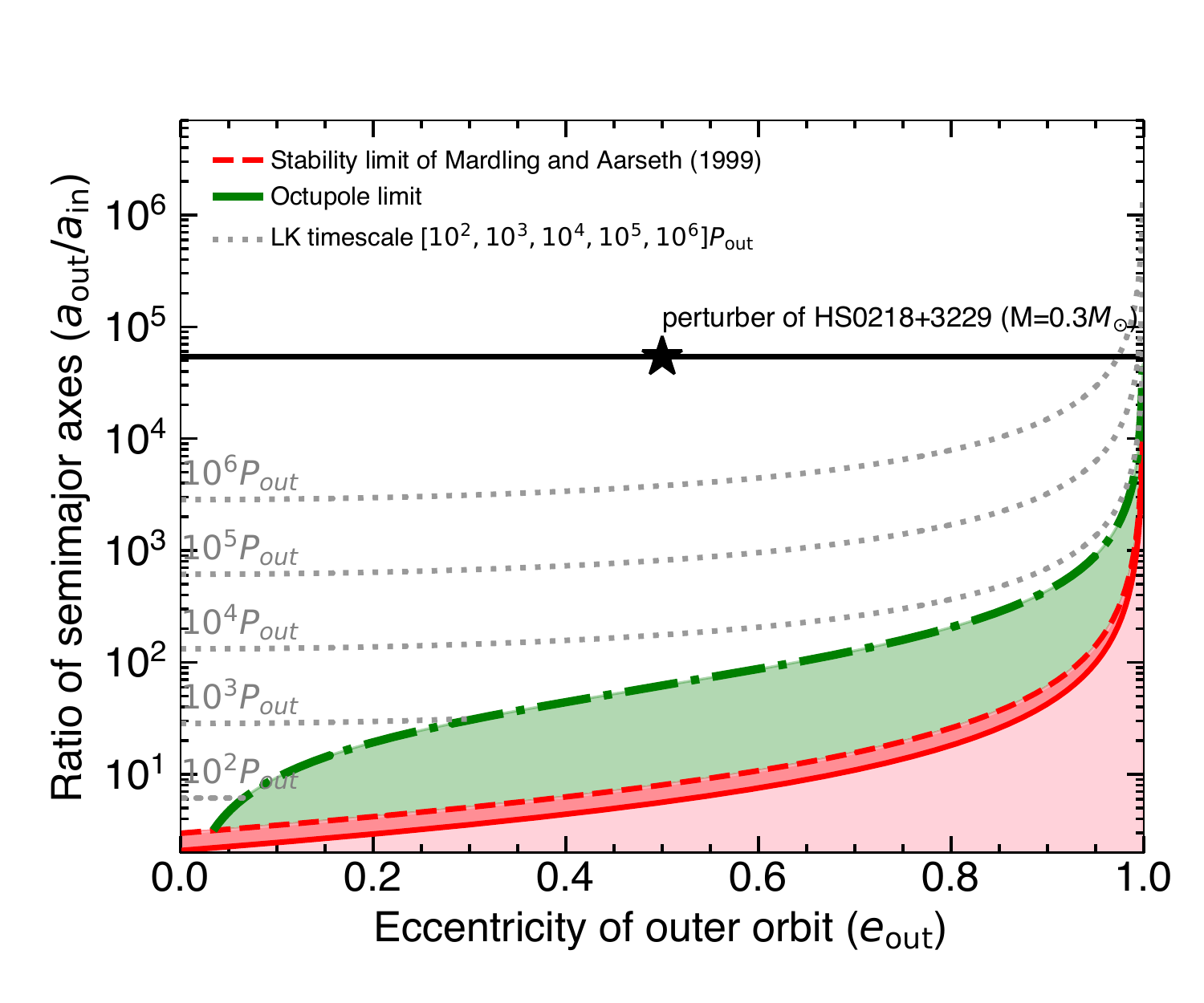}
    \caption{Dynamical regimes for a triple system consisting of an inner binary ($\Mwd=0.83\,\Msun, M_{\rm donor;i}=1.0\,\Msun$) and a third body with a mass of $0.5\,\Msun$ at 3000\,AU (star symbol). The black dotted lines are the constant Lidov-Kozai timescales (i.e. quadruple order) in units of the orbital period of the outer body. The octupole order term of the three-body interaction becomes important below the green area, where a limit in the range of $\epsilon_{\rm oct} = 0.01-0.001$ was assumed. The system is dynamically unstable below the red solid. \label{fig:kozai-lidov}}
\end{figure}

We used the stability criterion from \citet{Mardling-etal-1999ASIC..522..385M}, where the masses were taken from the initial binary configuration found by the MESA simulations (i.e. $\Mwd=0.83\,\Msun, M_{\rm donor;i}=1.0\,\Msun$), and a mass of 0.5\,\Msun\, for the third body was assumed. The stability limit is shown with red lines in Figure~\ref{fig:kozai-lidov} (inclination of 0º and 180º are demarcated by dashed and solid lines, respectively). The area below (pink shade) represents the unstable regime, which increases towards larger eccentricities of the outer body. No matter what the eccentricity of the outer body is, our triple system is dynamically stable.

However, it is known that an outer body can perturb the dynamics of the inner binary leading to long-term variations in the eccentricities and inclinations \citep{Kozai-1962AJ.....67..591K, Lidov-1962P&SS....9..719L}. The lowest level that is valid for an axisymmetric outer orbit (i.e. circular) is known as the standard Kozai mechanism (i.e. the quadrupole level of approximation). The timescales at which these oscillations are induced in inclination and eccentricity of the inner binary system are a function of the outer orbital period ($P_{\rm out}$) and are shown with dotted lines. Based on the projected separation, the outer period can be assumed as $P_{\rm out}\sim$110\,000 years. Thus \tar\ would need more than 10$^{11}$ years to manifest the oscillations due to the standard Kozai mechanism.

Generally, when |$\epsilon_{\rm oct}| \gtrsim 0.001-0.01$, the eccentric Lidov-Kozai mechanism (octupole level) can become important \citep{Naoz-etal-2011Natur.473..187N, Shappee-etal-2013ApJ...766...64S, Li-etal-2014ApJ...791...86L}. Hence, for completeness, the Eccentric Lidov-Kozai regime is shown in green (for $\epsilon_{\rm oct}=0.001-0.01$), which is negligible for the evolution of \tar.

Therefore we conclude that, if the system is indeed a hierarchical triple, it is possible that that tertiary could have influence to the formation of the CV, but not during the CV lifetime neither its future.

\section{Conclusions}
\label{sec:conclusions}

From the \textit{HST}/COS observations of \tar, 
we measured a radial velocity of the white dwarf of $K_{\mathrm{WD}}= 118 \pm 31$\,km\,s$^{-1}$, which agrees with previous estimates of \citet{Rodriguez-Gil-etal-2009A&A...496..805R}. In addition, we estimated the rotational velocity of the white dwarf to be $v~ {\rm sin}~i =
179 \pm 27$\,km\,s$^{-1}$. Ultraviolet spectroscopy can also provide accurate estimates of the white dwarf parameters, i.e. effective temperature, mass, radius, and abundances. However, the measured mass is very sensitive to the reddening and  higher resolution spectroscopy is needed to largely improve the abundance measurements. The carbon-to-nitrogen ratio extracted from \textit{HST}/COS ($\cno =-0.53^{+0.13}_{-0.14}$ and $\cno =-0.58^{+0.16}_{-0.15}$ for free and fixed reddening, respectively) has the potential to provide very strong constraints on the evolution of WD+eMS systems. Indeed, it is the most sensitive parameter, and behaves smoothly in the parameter space covered by our binary evolution simulations using MESA at the orbital periods of \tar. 

Our fits to a MESA grid, using GP processes, show that the initial mass of the donor is can be in the range of $0.90-0.98\,\Msun$, and the initial orbital period is $3.12-3.16$\,days when the white dwarf mass is assumed to be $\Mwd=0.60\,\Msun$. However, slightly lower values of orbital period ($2.88$\,days) and initial donor's mass ($0.91\,\Msun$) is needed when the white dwarf mass is fixed to $\Mwd=0.83\,\Msun$. We have found that \tar-like systems are not necessarily required to have undergone hydrogen burning in the past, as shown by our MESA simulations, not even for more extreme cases of mass ratio. Thus, \tar\ is not certain to have descended from a super soft X-ray source.  Neither is it mandatory to have experienced a phase of thermal timescale mass transfer as previously thought, which occurs if the initial mass ratio is $\gtrsim 1.5$. We expect that \tar\ will evolve into a system similar to EI\,Psc or V485-Cen, i.e. toward a system with an orbital period shorter than the minimum periods of CVs, and its spectrum will shows enhancement of helium but will still show hydrogen.

Systems of this type evolve very quickly, and in contrast to CVs, where the effective temperature decreases with time, in WD+eMS the effective temperature either remains roughly constant or increases toward shorter orbital periods. When the thermal timescale increases, below the period minimum, the effective temperature decreases again. 

Finally, the existence of a third object in orbit at 3000-3500\,AU has no effect on the evolution of \tar\ for timescales shorter than 10$^{11}$\,years, and thus it has no influence on its past or future evolution.

\section*{Acknowledgements}

OT was supported by a FONDECYT project 321038 and a Leverhulme Trust Research Project Grant. 
MRS and MZ were also supported by FONDECYT (grant 1221059).
The code from \href{https://github.com/TripleEvolution/dyn}{https://github.com/TripleEvolution/dyn} \citep{Toonen-etal-2020A&A...640A..16T} was adapted for Figure~\ref{fig:kozai-lidov}. 
We thank Patrick Godon for their helpful insights on the discussion. DdM is supported by INAF-ASI grants I/037/12/0 and 2017-14-H.0 and INAF/PRIN and "Main Streams" programmes. This project has received funding from the European Research Council
(ERC) under the European Union’s Horizon 2020 research and innovation programme (Grant agreement No. 101020057).


\section*{Data Availability}

The data and numerical tools used in this article can be obtained upon request to the corresponding author and after agreeing to the terms of use.



\bibliographystyle{mnras}
\bibliography{template} 




\appendix

\section{Some extra material}

it can be placed in an Appendix which appears after the list of references.

\subsection{Past evolution of \tar: Optimisation using Gaussian processes}

In this section, we describe in detail the methods used in Section \ref{sec:past_evolution} to estimate the values of the initial orbital period and mass of the donor that best describe the observed parameters of \tar. The method is equivalent for the two derived white dwarf masses.

The main goal is to find a point or a set of points $(P_{\mathrm{orb}}, M_{\mathrm{donor}})$ (initial orbital period and mass of the donor) that best describe the observed values of the parameters at an orbital period of 7.13h, denoted as \textit{optimal points}, even though we only know the value of the parameters at the points explored in the MESA grid. The parameters used are the accretion rate (log $\dot{M}$), the carbon-to-nitrogen ratio ($\cno$), the effective temperature ($T_{\mathrm{eff, donor}}$), and
the mass of the donor ($M_{\mathrm{donor}}$), marked with a star in Table \ref{tab:parameters}.

We applied two alternative methods where (1) a set of optimal points is found per parameter (Section \ref{sec:appendix_method1}) and (2) a single optimal point is found for all parameters based on an averaged weighted error (Section \ref{sec:appendix_method2}). Both methods use an interpolation method based on two-dimensional Gaussian processes (Section \ref{sec:appendix_gp}). The interpolation is constrained to orbital periods between $P_{\mathrm{orb,i}}=2.4-3.4$\,days and masses between $M_{\mathrm{donor}}=0.9-2.0M_{\mathrm{\odot}}$ for $\Mwd=0.83\,\Msun$, and orbital periods between $P_{\mathrm{orb,i}}=2.3-3.0$\,days and masses between $M_{\mathrm{donor}}=0.9-2.0M_{\mathrm{\odot}}$ for $\Mwd=0.60\,\Msun$.

\subsubsection{Method 1: Interpolation on the individual parameters}\label{sec:appendix_method1}
We implemented the following method for each parameter independently. For a given parameter, let $f(P_{\mathrm{orb}}, M_{\mathrm{donor}})$ be a function giving the value of the parameter at an orbital period 7.13h and any pair $(P_{\mathrm{orb}}, M_{\mathrm{donor}})$. Since we know $f$ only at the points in the MESA grid, we interpolate those values to estimate $f$ for any point $(P_{\mathrm{orb}}, M_{\mathrm{donor}})$ using a two-dimensional Gaussian process (described in Section \ref{sec:appendix_gp}). The interpolation provides the mean and marginal variance of the values of the parameter that would be observed at a point $(P_{\mathrm{orb}}, M_{\mathrm{donor}})$, distributed as a Gaussian distribution and denoted as $g_{(P_{\mathrm{orb}}, M_{\mathrm{donor}})}(p)$.

Let $h(p)$ be the empirical distribution of the observed value of the parameter. To find the set of optimal points that best describe the observed parameter, we choose the points $(P_{\mathrm{orb}}, M_{\mathrm{donor}})$ whose $g_{(P_{\mathrm{orb}}, M_{\mathrm{donor}})}(p)$ is \textit{similar} to the empirical distribution $h(p)$. Formally, $g$ and $h$ are \textit{similar} if 0 is between the 2.5 and 97.5 quantiles of $g - h$ ; that is, that the distribution of $g - h$ is centred around 0. 

This method provides a set of optimal points for each parameter, as shown in Figure \ref{fig:gp+error}.

\subsubsection{Method 2: Optimisation based on the averaged weighted error}\label{sec:appendix_method2}
The main purpose of this method is to find a single optimal pair $(P_{\mathrm{orb}}, M_{\mathrm{donor}})$ based on all parameters at once. First, we define an error measure for each pair $(P_{\mathrm{orb}}, M_{\mathrm{donor}})$ explored in the MESA grid to quantify how similar are the parameters of the track to the observed parameters of \tar. For a parameter $i$, let $x_i(P_{\mathrm{orb}}, M_{\mathrm{donor}})$ be the value of the parameter for the MESA evolutionary track at a point $(P_{\mathrm{orb}}, M_{\mathrm{donor}})$ and orbital period 7.13h, and let $h_i(p)$ be the empirical distribution of the observed values of the parameter $i$. The error of the track at $(P_{\mathrm{orb}}, M_{\mathrm{donor}})$ is given by the expected value of $|x_i - p|$ weighted by the probability of $p$ according to the empirical distribution $h$. That is:
\begin{equation}\label{eq:error_part_appendix}
\epsilon_i(P_{\mathrm{orb}}, M_{\mathrm{donor}}) = \int g(p)|x_i(P_{\mathrm{orb}}, M_{\mathrm{donor}}) - p| dp.
\end{equation}
Since $\epsilon_i$ is affected by the unit of measure of the parameter, we divide the error by the standard deviation of all tracks of the MESA grid, denoted by $\sigma_i$. Then, we average the error in (\ref{eq:error_part_appendix}) for all parameters; that is, the error measure is defined as:
\begin{equation}\label{eq:error_all_appendix}
\epsilon(P_{\mathrm{orb}}, M_{\mathrm{donor}}) = \frac{1}{4}\sum_{i=1}^4 \frac{\epsilon_i(P_{\mathrm{orb}}, M_{\mathrm{donor}})}{\sigma_i}.
\end{equation}

Second, we interpolate the error values of the MESA grid points to estimate the error measure at any pair $(P_{\mathrm{orb}}, M_{\mathrm{donor}})$ using a two-dimensional Gaussian process (described in Section \ref{sec:appendix_gp}). The interpolation provides the mean and marginal variance of the error that would be observed at a point $(P_{\mathrm{orb}}, M_{\mathrm{donor}})$, distributed as a Gaussian distribution. The mean obtained by the interpolation is denoted as $m(P_{\mathrm{orb}}, M_{\mathrm{donor}})$.

Finally, we apply the Nelder-Mead optimisation algorithm to find the point $(P_{\mathrm{orb}}, M_{\mathrm{donor}})$ that minimises mean $m$. We use the implementation of the optimiser found in the \textsc{Scipy v1.7.1} package in \textsc{Python} \citep{gao_implementing_2012}.

\subsubsection{Gaussian process fitting}\label{sec:appendix_gp}

The Gaussian processes are implemented using a Matérn covariance function with smoothness parameter $\nu=3/2$ and length scales $l_{\rm Porb}$ and $l_{\rm Mdonor}$. Covariance functions from the Matérn family are suitable for two-dimensional interpolation problems since the smoothness of the GP can be controlled by $\nu$ \citep{stein_interpolation_1999}. The model parameters are optimised by Maximum Likelihood using the L-BFGS-B algorithm with initial length scales $l_{\rm Porb}=0.5$\,days and $l_{\rm Mdonor}=0.5\,\Msun$. We use the \textsc{gpflow} package from \textsc{python} \citep{GPflow2017}.

\begin{figure*}
    \centering
    \includegraphics[width=18.cm]{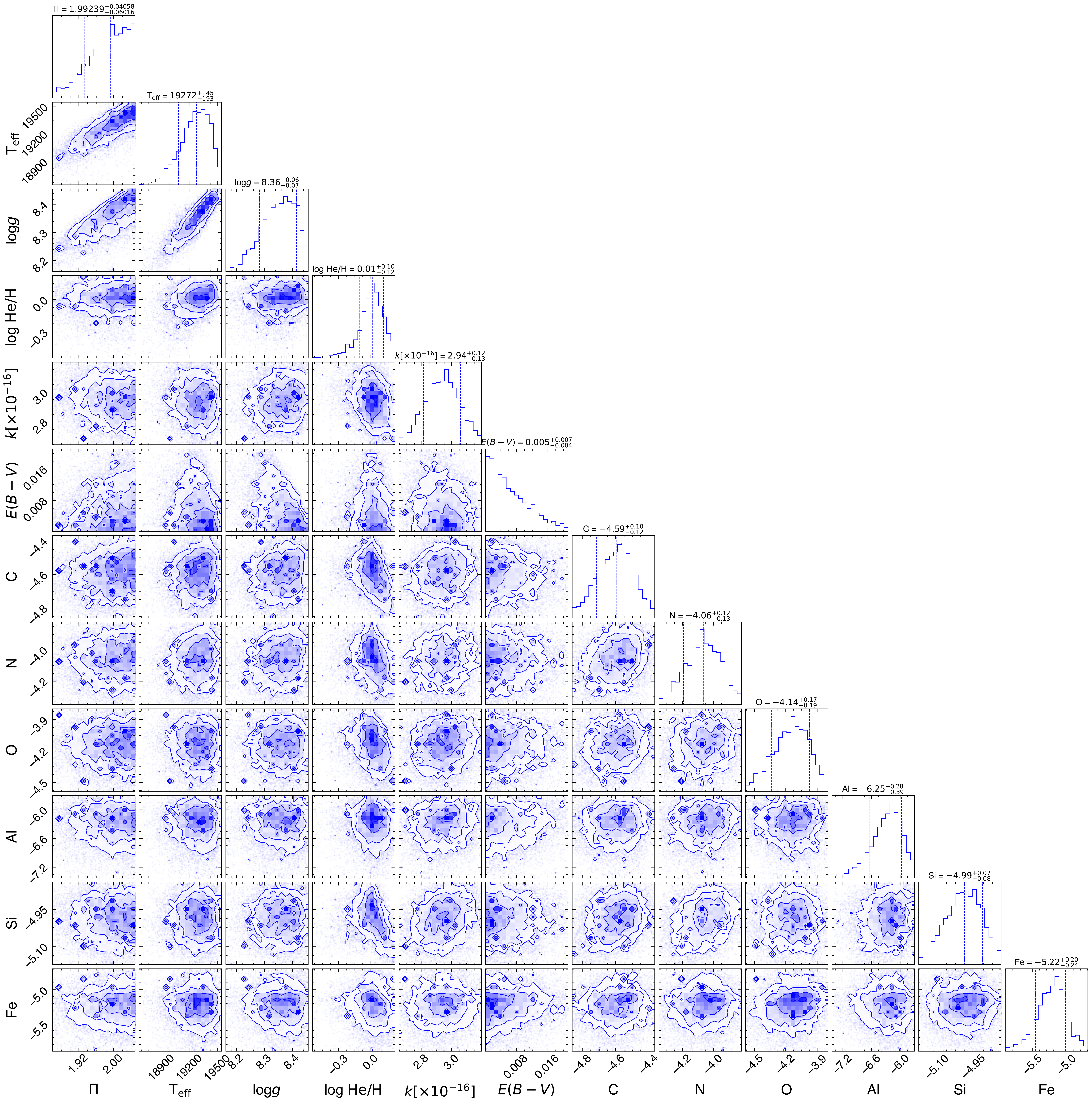}
    \caption{1-dimensional and 2-dimensional marginalized distribution of the samples for the MCMC fit to the COS spectrum with $E(B-V)$ as a free parameter. The best-fit value is taken as the median, while the errors are the 16$^{\mathrm{th}}$ and 84$^{\mathrm{th}}$ percentiles (dashed lines), which are shown on top of the 1-dimensional distributions. There is a strong correlation between the effective temperature (\Teff), the surface gravity (\logg) and the parallax (See section \ref{sec:results_fits} for details). In contrast, the chemical abundances do not present clear correlation among them. }
    \label{fig:corner}
\end{figure*}

\begin{figure*}
    \centering
    \includegraphics[width=18.cm]{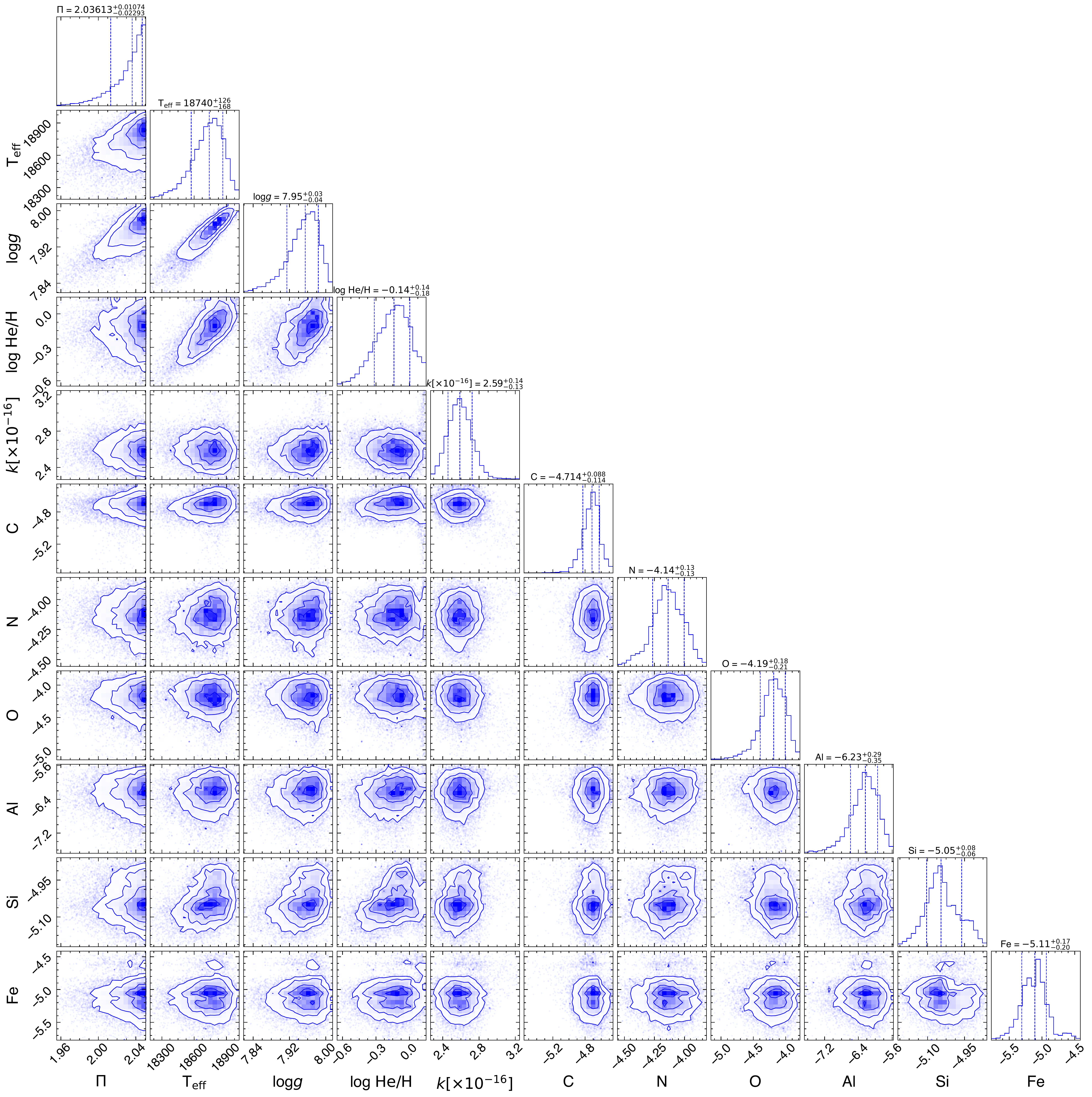}
    \caption{1-dimensional and 2-dimensional marginalized distribution of the samples for the MCMC fit to the COS spectrum with $E(B-V)=0.065$ fixed. The best-fit value is taken as the median, while the errors are the 16$^{\mathrm{th}}$ and 84$^{\mathrm{th}}$ percentiles (dashed lines), which are shown on top of the 1-dimensional distribution. There is a strong correlation between the effective temperature (\Teff), the surface gravity (\logg) and the parallax (See section \ref{sec:results_fits} for details). In contrast, the chemical abundances do not present clear correlation among them. }
    \label{fig:corner_ebv}
\end{figure*}


\bsp	
\label{lastpage}
\end{document}